\documentclass[conference]{IEEEtran}
\IEEEoverridecommandlockouts

\def\citetechreport{0}    

\usepackage{cite}
    \usepackage{amsthm}
\usepackage{amsmath,amssymb,amsfonts}
\usepackage{algorithmic}
\usepackage{graphicx}
\usepackage{textcomp}
\usepackage{xcolor}
\def\BibTeX{{\rm B\kern-.05em{\sc i\kern-.025em b}\kern-.08em
    T\kern-.1667em\lower.7ex\hbox{E}\kern-.125emX}}
    
\usepackage[font={footnotesize}]{subcaption}
\usepackage{algorithm,algorithmic}
\usepackage{mathtools}

\usepackage{color}

\allowdisplaybreaks

\usepackage{float}
\newfloat{algorithm}{t}{lop}

\usepackage{thmtools}

\declaretheoremstyle[%
 spaceabove=4pt,%
 spacebelow=4pt,%
 headfont=\normalfont\bfseries,%
 bodyfont=\normalfont\itshape,%
 postheadspace=0.5em,%
]{theoremstyle} 

\declaretheorem[name={Definition},style=theoremstyle]{definition}

\declaretheorem[name={Theorem},style=theoremstyle]{theorem}

\declaretheorem[name={Lemma},style=theoremstyle]{lemma}

\declaretheoremstyle[%
 spaceabove=0pt,%
 spacebelow=4pt,%
 headfont=\normalfont\itshape,%
 postheadspace=1em,%
 qed=\qedsymbol%
]{proofstyle} 

\declaretheorem[name={Proof},style=proofstyle,unnumbered]{prf}

\newcommand{\sz}[1]{s_{#1}}
\newcommand{\ca}[1]{c_{#1}}
\newcommand{\sr}{S}
\newcommand{\nd}{V}
\newcommand{\ur}{U}
\newcommand{\re}{\mathbb{R}^+}
\newcommand{\ui}[1]{\xi_{#1}}
\newcommand{\us}[1]{W_{#1}}
\newcommand{\uw}[1]{r_{#1}}
\newcommand{\li}{i}
\newcommand{\lj}{j}
\newcommand{\lk}{k}

\newcommand{\se}[1]{X_{#1}}
\newcommand{\rp}{X}
\newcommand{\id}[1]{\mathcal{I}\left(#1\right)}

\newcommand{\lp}[2]{\omega_{#1, #2}}
\newcommand{\lr}[1]{\alpha_{#1}}
\newcommand{\pa}[2]{\Omega_{#1, #2}}
\newcommand{\ld}{q}

\newcommand{\ns}[2]{\eta_{#1, #2}}
\newcommand{\su}{\sigma}
\newcommand{\sv}[1]{\mu(#1)}
\newcommand{\sw}[1]{\lambda(#1)}

\newcommand{\ls}{\Lambda}

\newcommand{\pr}[1]{\mathbb{P}\left(#1\right)}
\newcommand{\rw}{R}
\newcommand{\arw}{\hat{R}}

\newcommand{\fb}{\beta}
\newcommand{\fc}{\gamma}
\newcommand{\fd}{\delta}

\newcommand{\de}[2]{\Delta_{#1, #2}}
\newcommand{\sla}{\tau}
\newcommand{\psa}{\tau'}
\newcommand{\sfa}[1]{\mathcal{T}_{#1}}
\newcommand{\sat}[2]{f_{#1, #2}}
\newcommand{\ex}[2]{#1^*_{#2}}

\newcommand{\rpt}[1]{X^{#1}}
\newcommand{\set}[2]{X^{#1}_{#2}}
\newcommand{\slb}[2]{q_{#1, #2}}

\newcommand{\exr}[1]{\mathcal{E}\left(#1\right)}
\newcommand{\fsa}{\psi}
\newcommand{\fm}{\lambda}
\newcommand{\vo}[1]{\delta'_{#1}}
\newcommand{\pf}[1]{\theta_{#1}}
\newcommand{\pg}[1]{\theta'_{#1}}

\newcommand{\pq}[1]{Q_{#1}}
\newcommand{\ha}{1}
\newcommand{\hb}{2}
\newcommand{\hc}{3}

\newcommand{\bsl}{\oplus}
\newcommand{\ssl}{\ominus}
\newcommand{\map}{\zeta}
\newcommand{\pap}{\zeta'}

\newcommand{\mas}[1]{\mathcal{Y}_{#1}}
\newcommand{\exs}[1]{\mathcal{D}(#1)}
\newcommand{\pbl}[3]{\epsilon_{#1,#2,#3}}

\newcommand{\ag}{U}
\newcommand{\lel}{l}
\newcommand{\agf}[1]{\hat{r}_{#1}}

\newcommand{\eve}[2]{E_{#1,#2}}

\begin{document}

\title{Service Placement with Provable Guarantees in Heterogeneous Edge Computing Systems \vspace{-0.1in}}

\author{
\IEEEauthorblockN{Stephen Pasteris\IEEEauthorrefmark{1}, Shiqiang Wang\IEEEauthorrefmark{2}, Mark Herbster\IEEEauthorrefmark{1}, Ting He\IEEEauthorrefmark{3}}
\IEEEauthorblockA{\IEEEauthorrefmark{1}University College London, London, UK. Email: \{s.pasteris, m.herbster\}@cs.ucl.ac.uk}
\IEEEauthorblockA{\IEEEauthorrefmark{2}IBM T. J. Watson Research Center, Yorktown Heights, NY, USA. Email: wangshiq@us.ibm.com}
\IEEEauthorblockA{\IEEEauthorrefmark{3}Pennsylvania State University, University Park, PA, USA. Email: t.he@cse.psu.edu}
\vspace{-0.3in}
\thanks{This is an extended version of the paper with the same title presented at IEEE INFOCOM 2019.

This research was sponsored by the U.S. Army Research Laboratory and the U.K. Ministry of Defence under Agreement Number W911NF-16-3-0001. The views and conclusions contained in this document are those of the authors and should not be interpreted as representing the official policies, either expressed or implied, of the U.S. Army Research Laboratory, the U.S. Government, the U.K. Ministry of Defence or the U.K. Government. The U.S. and U.K. Governments are authorized to reproduce and distribute reprints for Government purposes notwithstanding any copyright notation hereon.}
}

\maketitle

\begin{abstract}
Mobile edge computing (MEC) is a promising technique for providing low-latency access to services at the network edge. The services are hosted at various types of edge nodes with both computation and communication capabilities. Due to the heterogeneity of edge node characteristics and user locations, the performance of MEC varies depending on where the service is hosted. In this paper, we consider such a heterogeneous MEC system, and focus on the problem of placing multiple services in the system to maximize the total reward. We show that the problem is NP-hard via reduction from the set cover problem, and propose a deterministic approximation algorithm to solve the problem, which has an approximation ratio that is not worse than $\left(1-e^{-1}\right)/4$. The proposed algorithm is based on two sub-routines that are suitable for small and arbitrarily sized services, respectively. The algorithm is designed using a novel way of partitioning each edge node into multiple slots, where each slot contains one service. The approximation guarantee is obtained via a specialization of the method of conditional expectations, which uses a randomized procedure as an intermediate step. In addition to theoretical guarantees, simulation results also show that the proposed algorithm outperforms other state-of-the-art approaches.
\end{abstract}


\section{Introduction}
\label{sec:introduction}

\begin{figure}
\centering
\includegraphics[width=0.85\columnwidth]{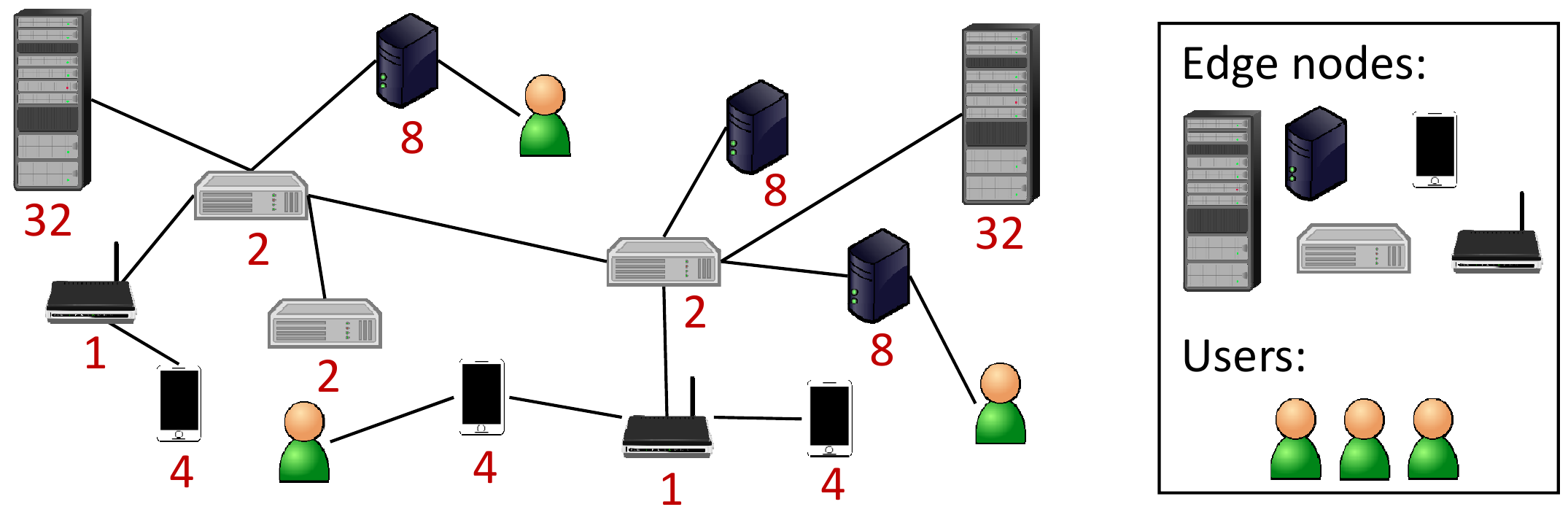}
\caption{Example of an MEC system with heterogeneous edge nodes that are arbitrarily interconnected with each other. The numbers below each edge node indicate some capacity notion (such as storage capacity for service programs) of the node. The user can connect to the system via any edge node.}
\label{fig:architecture}
\end{figure}

Many emerging applications such as the Internet of Things (IoT), virtual/augmented reality, etc. require low-latency access to services at the network edge.
Mobile edge computing (MEC) has emerged as a key technology to make this possible~\cite{MachMECSurvey,MaoMECSurvey}. In MEC, services are hosted at edge nodes with communication, computation, and storage capabilities. The edge nodes can include micro servers, IoT gateways, routers, mobile devices, etc. They are connected to the wide-area network (WAN) and provide low-latency service to users that are within a suitable communication distance.
An example of an MEC system is shown in Fig.~\ref{fig:architecture}.

A major challenge in MEC is to decide which services each edge node should host in order to satisfy the user demand, which we refer to as the \emph{service placement} problem. The service placement has to take into account the heterogeneity of edge nodes, services, and users. For example, the response time of edge services can vary significantly depending on the network interface and hardware configuration of edge nodes~\cite{chen2017empirical}. Users can have different communication latencies to different edge nodes.
Different services may consume different amount of resource and are compatible with different operating systems and hardware. 
All these aspects pose significant challenges to solving the service placement problem.

Due to these challenges, existing work on service placement often has limitations in terms of practicality and performance guarantee. Heuristic algorithms without approximation guarantees are proposed in~\cite{TongHierarchical2016,Ceselli2017,Xu2016,Jia2017}.
Among those approaches that provide approximation/optimality guarantees, \cite{ChenOffload2016,TongOffloading2016,XiaoOffloading2017,HouJSAC2018} focus on offloading decisions, which do not consider the placement of services onto multiple edge nodes that can host services for other users.
The work in~\cite{TanInfocom2017} focuses on job scheduling for multiple edge nodes, but does not incorporate parallel execution of multiple services at the same edge node.
Others consider the trade-off between delay and energy~\cite{urgaonkar2015performance,xu2018joint}, which neglect the heterogeneity of edge node platforms where some service may require hardware that only exists on some specific edge nodes. 
Elastic services that can be partitioned in arbitrary ways is considered in~\cite{WangICDCS2017}, which can be unrealistic in practice because it is usually impossible to split a computer program arbitrarily. The work in~\cite{Wang16TPDS,WangOnlinePlacement2017,wang2018service} does not consider concrete capacity limits, which therefore cannot capture the strict resource limitation of edge nodes. The work in~\cite{TingICDCS2018} proposes a greedy service placement algorithm that can be shown to have a constant approximation ratio when all services have the same size and the reward is homogeneous. No approximation guarantee is shown for the heterogeneous setting. 
In addition, many of the existing algorithms, other than~\cite{TingICDCS2018}, only guarantee non-constant approximation ratios.

Different from the above existing work, in this paper, we consider an MEC system with 1) non-splittable service entities that are shareable among multiple users, 2) heterogeneous service and edge node sizes, 3) heterogeneous rewards of serving users, and 4) strict capacity limits of edge nodes. There are an arbitrary number of edge nodes, services, and users, where each user requires a service. We propose a \emph{constant factor approximation algorithm} to find a feasible service placement that maximizes the total system reward. 

The challenge in our problem is that many standard  techniques for approximation algorithms, such as those used in~\cite{TingICDCS2018}, are either not applicable or can only provide a bad approximation ratio. For example, as we will show in Section~\ref{subsec:GreedyBadExample}, the greedy algorithm used in~\cite{TingICDCS2018} can perform arbitrarily badly for our problem.
We design our algorithm based on a novel approach that partitions each node into multiple slots, together with a highly non-trivial way of applying the idea of the method of conditional expectations~\cite{vazirani2013approximation}.

Our main contributions in this paper are as follows:

1) We formulate the general service placement (GSP) problem as described above, and convert it to an equivalent problem that we call service placement with set constraints (SPSC) which is easier to approximate. 
We show that both GSP and SPSC are NP-hard.

2) For the case where all services are small compared to the capacities of edge nodes, we propose an algorithm that solves SPSC with an approximation ratio\footnote{For the maximization problem we consider in this paper, we define the approximation ratio $\rho$ as $\rho \cdot \mathrm{OPT} \leq R^* \leq \mathrm{OPT}$, where $R^*$ is the solution from the approximation algorithm and $\mathrm{OPT}$ is the true optimal solution. A larger approximation ratio indicates a better performance.} of $1-e^{-\left(1-\sqrt{\fb}\right)^2}$, where $\fb$ is the maximum size of any service divided by the minimum capacity of any node.

3) For the general case with arbitrary service sizes and node capacities, we propose an algorithm that solves SPSC with an approximation ratio of $\left(1-e^{-1}\right) / 4$.

4) We combine the above two algorithms and propose an algorithm that works for the general case and has an approximation ratio of $\max\left\{ 1-e^{-\left(1-\sqrt{\fb}\right)^2} ; \left(1-e^{-1}\right) / 4 \right\}$.

5) We present simulation results that show the effectiveness of our proposed algorithms empirically.

\section{Problem Formulation}

\subsection{General Service Placement (GSP)}

In this paper, we aim at solving the GSP defined as follows.

\subsubsection{Input} \label{subsec:GSPInput}
Let $\sr$ denote the set of services, $\nd$ denote the set of nodes, and $\ag$ denote the set of users. 
For all $\li\in\sr$, we define some $\sz{\li}\in\re$ as the size of service $\li$.
For all $\lj\in\nd$, we define some $\ca{\lj}\in\re$ as the capacity of node $\lj$.
For all $\lk\in\ag$, we define some $\ui{\lk}\in\sr$ as the service required by user $\lk$.
For all $\lk\in\ag$, we define a map $\agf{\lk}:\nd\rightarrow\mathbb{R}^+$, where $\agf{\lk}(\lj)$ is the reward obtained for serving user $\lk$ if its service $\ui{\lk}$ is provided by node $\lj$. The size and capacity definitions can be related to storage or any other type of resource that can be shared among multiple users. Users that require non-shareable resource can be considered as requiring different services.
The reward can be defined as related to the service quality (such as response time) perceived by the user, and its value can differ by users, services, and service placement configurations. 
Such a reward definition can capture the heterogeneity in the operation system, hardware, and networking aspects.
We also note that a user does not need to be a ``real'' user, it can be any instance of service request.

\subsubsection{Service Placement}
A \textit{service placement} $\rp$ is an indexed set $\{\se{\li}:\li\in\sr\}$ where, for every $\li\in\sr$, we have $\se{\li}\subseteq\nd$, and $\se{\li}$ is the set of nodes that service $\li$ is placed on. A service placement $\rp$ is \textit{feasible} if and only if (iff): 
\begin{equation}
\sum_{\li}\sz{\li}\cdot\id{\lj\in\se{i}} \leq \ca{\lj}
\label{eq:feasibilityGSP}
\end{equation}
for all $\lj\in\nd$, where we define $\id{\cdot}$ to be the indicator function, i.e., $\id{E}:=1$ if $E$ is true and $\id{E}:=0$ otherwise. Condition (\ref{eq:feasibilityGSP}) states that the total size of services hosted at a node does not exceed its capacity.

\subsubsection{Objective}
In our system, if there are multiple nodes containing service $\ui{\lk}$, then the service of user $\lk$ is provided by a node $\lj$ that gives the maximum reward $\agf{\lk}(\lj)$. If no node contains $\ui{\lk}$ then we obtain no reward from user $\lk$.
With this definition, given a service placement $\rp$, the reward given to us by a user $\lk$ is equal to $\max_{\lj\in\se{\ui{\lk}}}\agf{\lk}(\lj)$, where $\max_{\lj\in\emptyset}\agf{\lk}(\lj):=0$. 
The objective of GSP to find a feasible service placement $\rp$ that  maximizes the total reward, i.e.: 
\begin{align}
\max_{\rp} \quad & \sum_{\lk\in\ag}\max_{\lj\in\se{\ui{\lk}}}\agf{\lk}(\lj) \\
\mathrm{s.t.} \quad & \textrm{Condition (\ref{eq:feasibilityGSP})}. \nonumber
\end{align}

\subsection{Service Placement with Set Constraints (SPSC)}
\label{subsec:SPSCIntro}

We first introduce the SPSC problem as follows. Then, we will show that both GSP and SPSC are NP-hard. Afterwards, we show how to convert any GSP to an equivalent SPSC and focus on approximation algorithms for SPSC.

In SPSC, we reuse the definitions in Section~\ref{subsec:GSPInput} except for the reward. The reward for each user in SPSC is defined by a subset of nodes as follows.
For all $\lk\in\ur$, we define some $\us{\lk}\subseteq\nd$ to represent a set of nodes, one of these nodes must contain service $\ui{\lk}$ in order for user $\lk$ to be satisfied.
For all $\lk\in\ur$, we define some $\uw{\lk}\in\re$ to represent the reward received by the system if user $\lk$ is satisfied.

Given a service placement $\rp$, a user $\lk$ is satisfied (so the system receives reward $\uw{\lk}$) iff $\us{\lk}\cap\se{\ui{\lk}}\neq\emptyset$, i.e., service $\ui{\lk}$ is placed on some node in $\us{\lk}$. Our objective is to find a feasible service placement that maximizes the total reward: 
\begin{align}
\max_{\rp} \quad & \sum_{\lk\in\ur}\uw{\lk}\cdot\id{\us{\lk}\cap\se{\ui{\lk}}\neq\emptyset} \label{eq:SPSCProblem}\\
\mathrm{s.t.} \quad & \textrm{Condition (\ref{eq:feasibilityGSP})} . \nonumber
\end{align}
Let $\rw$ denote the true (but unknown) optimal value of the objective in (\ref{eq:SPSCProblem}).

\begin{theorem}
\label{theorem:NPHardnessGSP}
Both GSP and SPSC are NP-hard.
\end{theorem}
\begin{prf}
The proof is based on reduction from the decision version of the set cover problem, which is NP-complete~\cite{karp1972reducibility}.
\if\citetechreport1
    See our technical report~\cite[Appendix~\ref{append:NPHardProof}]{techReportThisPaper} for details.
\else
    See Appendix~\ref{append:NPHardProof} for details.
\fi
\end{prf}

\subsection{Converting GSP to SPSC}

Because GSP is NP-hard, we seek for approximate solutions. It is difficult to approximate GSP directly. Therefore, we transform GSP to an equivalent SPSC problem, and propose approximation algorithms for SPSC.

Suppose we have a GSP instance. We will now construct an equivalent instance of SPSC. For clarity we will, in this subsection, refer to the users we construct for the SPSC instance as ``restricted users'' and define $\ur'$ to be the set of restricted users we construct. Note that $\ur$ is the set of users in the original GSP instance. The sets $\sr$ and $\nd$, as well as the values $\{\sz{\li}:\li\in\sr\}$ and $\{\ca{\lj}:\lj\in\nd\}$ in the SPSC instance are the same as in the GSP instance. This implies that a service placement is feasible for the SPSC instance iff it is feasible for the GSP instance.

\begin{algorithm}
\caption{Converting GSP to SPSC}
\label{AN}
\begin{algorithmic}[1]
  \footnotesize
  \STATE For all $\lel\in\ag$:
  \STATE\label{orderline}~~ Order $\nd$ as $\lj_{\lel, 1}, \lj_{\lel, 2},... , \lj_{\lel, |\nd|}$ so that $\agf{\lel}(\lj_{\lel, b})\!\geq\! \agf{\lel}(\lj_{\lel, b+1}),\forall b \! <\! |\nd|$;
  \STATE ~~~~~~~ For all $b\leq |\nd|$:
  \STATE~~~~~~~~~~~~~~ Create a restricted user $\lk_{(\lel, b)}$ with:
  \STATE~~~~~~~~~~~~~~~~~~~~~ $\ui{\lk_{(\lel, b)}}\leftarrow\ui{\lel}$;
  \STATE\label{line6}~~~~~~~~~~~~~~~~~~~~~ $\us{\lk_{(\lel, b)}}\leftarrow\{\lj_{\lel, b'}:b'\leq b\}$;
  \STATE\label{line7}~~~~~~~~~~~~~~~~~~~~~ If $b<|\nd|$ then  $\uw{\lk_{(\lel, b)}}\leftarrow\agf{\lel}(\lj_{\lel, b})-\agf{\lel}(\lj_{\lel, b+1})$;
  \STATE\label{line8}~~~~~~~~~~~~~~~~~~~~~ If $b=|\nd|$ then $\uw{\lk_{(\lel, b)}}\leftarrow\agf{\lel}(\lj_{\lel, b})$;
  \STATE Output $\ur'\leftarrow\{\lk_{(\lel, b)}:\lel\in\ag, b\leq|\nd|\}$;
\end{algorithmic}
\end{algorithm}

The set of restricted users, as well as their associated sets, services and rewards, is constructed in Algorithm \ref{AN}. In the rest of this subsection, we will use the notation introduced in Algorithm \ref{AN} to show that the two problems are equivalent. 

\begin{theorem}\label{redth}
If we have a feasible service placement $\rp$, then for all $\lel\in\ag$ we have:
$$\max_{\lj\in\se{\ui{\lel}}}\agf{\lel}(\lj)=\sum_{b\leq|\nd|}\uw{\lk_{(\lel, b)}}\cdot\id{\us{\lk_{(\lel,b)}}\cap\se{\ui{\lk_{(\lel, b)}}}\neq\emptyset}.$$
\end{theorem}

\begin{prf}

Define $\hat{b}:=\min\{b:\lj_{\lel, b}\in \se{\ui{\lel}}\}$.
For all $b'<\hat{b}$ we have $\lj_{\lel, b'}\notin \se{\ui{\lel}}$. Also for all $b<\hat{b}$ we have, by Line \ref{line6} of Algorithm~\ref{AN}, that  $\us{\lk_{(\lel, b)}}\subseteq\{\lj_{b'}:b'<\hat{b}\}$ which hence does not intersect with $\se{\ui{\lel}}$.  
On the other hand, if $b\geq \hat{b}$ then, by Line \ref{line6} of Algorithm~\ref{AN}, we have $\lj_{\lel, \hat{b}}\in\us{\lk_{(\lel, b)}}$. By definition $\lj_{\lel, \hat{b}}$ is also in $\se{\ui{\lel}}$ so $\us{\lk_{(\lel, b)}}$ intersects with $\se{\ui{\lel}}$.

Since for all $b\leq|\nd|$ we have  $\se{\ui{\lk_{(\lel, b)}}}=\se{\ui{\lel}}$, we have shown that for any $b\leq|\nd|$ we have $\us{\lk_{(\lel, b)}}\cap\se{\ui{\lk_{(\lel, b)}}}\neq\emptyset$ if and only if $b\geq \hat{b}$. This implies:
{\footnotesize
\begin{align*}
&\sum_{b\leq|\nd|}\uw{\lk_{(\lel, b)}}\cdot\id{\us{\lk_{(\lel,b)}}\cap\se{\ui{\lk_{(\lel, b)}}}\neq\emptyset}
= \sum_{\hat{b}\leq b\leq|\nd|}\uw{\lk_{(\lel, b)}}\\
& =\agf{\lel}(\lj_{\lel, |\nd|})+\sum_{\hat{b}\leq b<|\nd|}(\agf{\lel}(\lj_{\lel, b})-\agf{\lel}(\lj_{\lel, b+1}))
=\agf{\lel}(\lj_{\hat{b}}).
\end{align*} }%
where the second equality follows from Lines \ref{line7} and \ref{line8} of Algorithm~\ref{AN}. Now suppose there exists some node $\lj'\in\nd$ which satisfies $\agf{\lel}(\lj')>\agf{\lel}(\lj_{\hat{b}})$. Let $b'$ be such that $\lj'=\lj_{\lel, b'}$. By the ordering in Line \ref{orderline} we then have $b'<\hat{b}$ and hence $\lj_{\lel, b'}\notin\se{\ui{\lel}}$. This shows that $\max_{\lj\in\se{\ui{\lel}}}\agf{\lel}(\lj)=\agf{\lel}(\lj_{\hat{b}})$ (as $\lj_{\hat{b}}\in\se{\ui{\lel}}$) which, combining with above, proves the theorem.
\end{prf}

By Theorem \ref{redth} we have:
{\footnotesize
\begin{align*}
\sum_{\lel\in\ag}\max_{\lj\in\se{\ui{\lel}}}\agf{\lel}(\lj) 
&= \sum_{\lel\in\ag}\sum_{b\leq|\nd|}\uw{\lk_{(\lel, b)}}\cdot\id{\us{\lk_{(\lel,b)}}\cap\se{\ui{\lk_{(\lel, b)}}}\neq\emptyset}\\
&= \sum_{\lk\in\ur'}\uw{\lk}\cdot\id{\us{\lk}\cap\se{\ui{\lk}}\neq\emptyset}.
\end{align*} }%
Thus, for any feasible\footnote{Note that a service placement $\rp$ that is feasible for GSP is also feasible for SPSC, and vice versa, because the feasibility of both GSP and SPSC are specified by (\ref{eq:feasibilityGSP}).} service placement $\rp$, the reward of the GSP instance is the same as the reward of the SPSC instance. 
This shows that with the conversion given in Algorithm~\ref{AN}, the two problems are equivalent.

\textbf{Outline:} In the following, we present algorithms to solve SPSC with approximation guarantees.
The approximation algorithm starts with solving a linear program (LP) presented in Section~\ref{sec:LinearProgrammingStep}.
Then, the algorithm is based on a notion of ``slot allocation'' that will be explained later. 
We present two slot allocation algorithms, referred to as SA1 and SA2, in Sections~\ref{sec:SA1} and \ref{sec:SA2}, respectively. 
Then, in Section~\ref{sec:OverallAlgorithm}, we present an algorithm that combines SA1 and SA2 which is the final algorithm for solving SPSC (and thus GSP).
Section~\ref{sec:Simulation} presents simulation results. Section~\ref{sec:RelatedWork} discusses some further related work, and Section~\ref{sec:Conclusion} draws conclusion.

\section{Linear Programming Step}
\label{sec:LinearProgrammingStep}

Define indexed sets $\{\lp{\li}{\lj}:\li\in\sr,\lj\in\nd\}\subseteq\re$ and $\{\lr{\lk}:\lk\in\ur\}\subseteq\re$.
Both SA1 and SA2 for solving SPSC solve the following LP as a first step:
\begin{subequations}  \label{eq:LinearProgram}
\begin{align}
\max_{\{\lp{\li}{\lj}\}, \{\lr{\lk}\}} \quad & \arw := \sum_{\lk\in\ur}\lr{\lk}\uw{\lk}  \\
\mathrm{s.t.} \quad &  \lr{\lk}\leq\!\sum_{\lj\in\us{\lk}}\lp{\ui{\lk}}{\lj}, \!\! && \forall \lk\in\ur, \\
& \lr{\lk}\leq 1,\!\!  && \forall \lk\in\ur, \\
& \sum_{\li\in\sr}\lp{\li}{\lj}\sz{\li}\leq\ca{\lj}, \!\!&& \forall \lj\in\nd, \\
& \lp{\li}{\lj}=0,\!\! && \forall \li\in\sr, \lj\in\nd : \sz{\li}>\ca{\lj} , \\
& 0 \leq \lp{\li}{\lj}\leq 1,\!\! && \forall \li\in\sr, \lj\in\nd .
\end{align}
\end{subequations}

\begin{theorem}\label{BOR}
We have $\arw\geq\rw$. 
\end{theorem}

\begin{prf}
Choose a feasible service placement $\rp$ that has a total reward of $\rw$. If we then define $\lp{\li}{\lj}:=\id{\lj\in\se{\li}}$ and $\lr{\lk}:=\id{\us{\lk}\cap\se{\ui{\lk}}\neq\emptyset}$ it is clear that all the above constraints are satisfied and $\sum_{\lk\in\ur}\lr{\lk}\uw{\lk}=\rw$. Hence, if we choose $\{\lp{\li}{\lj}:\li\in\sr,\lj\in\nd\}\subseteq\re$ and $\{\lr{\lk}:\lk\in\ur\}\subseteq\re$ that satisfy the constraints and maximize $\sum_{\lk\in\ur}\lr{\lk}\uw{\lk}$ we must have $\sum_{\lk\in\ur}\lr{\lk}\uw{\lk}\geq\rw$.
\end{prf}

\section{First Slot Allocation Algorithm (SA1)}
\label{sec:SA1}

For both SA1 and SA2, we define:
\begin{align}
\fc  :=1-\sqrt{\fb};\quad\quad \fd  :=(1-\sqrt{\fb})^2  \label{eq:defGammaDelta}
\end{align}
for some given $\fb < 1$. We use $\mathbb{N}$ to denote the set of natural numbers (excluding zero) throughout the paper.

Our first algorithm, SA1, for solving SPSC is used in the case that we have $\max_{\li\in\sr}\sz{\li} \leq\fb(\min_{\lj\in\nd}\ca{\lj})$ for some given $\fb<1$. SA1 has an approximation ratio of $1-e^{-\left(1-\sqrt{\fb}\right)^2}$, which will be shown in Theorem~\ref{theorem:SA1ApproxRatio} later.

\begin{definition}
\label{def:SA1}
After solving the LP in (\ref{eq:LinearProgram}) we define the following for SA1. $\forall\lj\in\nd, \ld\in\mathbb{N}$:

\vspace{-0.13in}
{\footnotesize
\begin{align*}
\pa{\lj}{\ld}&:=\{\li\in\sr:\fc^{\ld}\ca{\lj}\fb\!<\!\sz{\li}\!\leq\!\fc^{\ld-1}\ca{\lj}\fb\} ; \!
&&\de{\lj}{\ld}:=\sum_{\li\in\pa{\lj}{\ld}}\lp{\li}{\lj} ;\\
\vo{\lj}&:=\frac{\fd\ca{\lj}}{\sum_{\li\in\sr}\sz{\li}\lp{\li}{\lj}} ;
&&\ns{\lj}{\ld}:=\left\lceil\vo{\lj}\de{\lj}{\ld}\right\rceil.
\end{align*} }%
\end{definition}

\begin{algorithm}
\caption{Slot Creation of SA1}
\label{A1}
\begin{algorithmic}[1]
  \footnotesize
  \STATE For all $\lj\in\nd$ and $\ld\in\mathbb{N}$ such that $\pa{\lj}{\ld}\neq\emptyset$:
  \STATE~~~~~~~ Create $\ns{\lj}{\ld}$ slots $\su$ with $\sv{\su}\leftarrow\lj$ and $\sw{\su}\leftarrow\ld$;
  \STATE Output $\ls$ as the set of all slots created;
\end{algorithmic}
\end{algorithm}

\begin{algorithm}
\caption{Service Placement}
\label{A2}
\begin{algorithmic}[1]
  \footnotesize
  \STATE Receive $\ls$ from the slot creation algorithm;
  \STATE For every $\su\in\ls$: set $\psa(\su)\leftarrow\emptyset$;
  \STATE For every $\su\in\ls$:
  \STATE \label {lia}\label{fs} ~~~~~~~  For every $\li\in\pa{\sv{\su}}{\sw{\su}}$:
  \STATE ~~~~~~~~~~~~~~ $\ex{\sla}{\li}(\su)\leftarrow i$;
  \STATE ~~~~~~~~~~~~~~ For all $\su'\setminus\{\su\}$: set $\ex{\sla}{\li}(\su')\leftarrow \sla(\su')$;
  \STATE ~~~~~~~ $\li'\leftarrow\operatorname{argmax}_{\li\in\pa{\sv{\su}}{\sw{\su}}}\exr{\ex{\sla}{\li}}$;  \label{alg:ServicePlacement:computeMaxExpect}
  \STATE \label{lib}~~~~~~~ $\psa(\su)\leftarrow\li'$;
  \STATE $\fsa\leftarrow\psa$;
  \STATE For all $\li\in\sr$:
  \STATE ~~~~~~~ Output $\set{\fsa}{\li}\leftarrow\{\lj\in\nd:\exists\su\operatorname{~with~}\sv{\su}=\lj, \sla(\su)=\li\}$;
\end{algorithmic}
\end{algorithm}

\begin{algorithm}
\caption{Computing $\exr{\ex{\sla}{}}$}
\label{A3}
\begin{algorithmic}[1]
  \footnotesize
  \STATE Receive $\ls$ from the slot creation algorithm;
  \STATE For all $\lk\in\ur$ such that $\exists\su\in\ls$ with $\sv{\su}\in\us{\lk}, \ex{\sla}{}(\su)=\ui{\lk}$:
  \STATE ~~~~~~~$\pf{\lk}\leftarrow1$;
  \STATE For all $\lk\in\ur$ such that $\not\exists\su\in\ls$ with $\sv{\su}\in\us{\lk}, \ex{\sla}{}(\su)=\ui{\lk}$:
  \STATE\label{begp}~~~~~~~$p\leftarrow 1$;
  \STATE~~~~~~~For all $\su\in\ls$ with $\sv{\su}\in\us{\lk}$, $\ex{\sla}{}(\su)=\emptyset$, $\ui{\lk}\in\pa{\sv{\su}}{\sw{\su}}$:
  \STATE~~~~~~~~~~~~~~$p\leftarrow \left(1-\lp{\ui{\lk}}{\sv{\su}}/\de{\sv{\su}}{\sw{\su}}\right)p$;
  \STATE\label{endp}~~~~~~~$\pf{\lk}=1-p$;
  \STATE Output $\exr{\ex{\sla}{}}\leftarrow\sum_{\lk\in\ur}\pf{\lk}\uw{\lk}$;
\end{algorithmic}
\end{algorithm}

\textbf{Procedure of SA1:} First, we solve the LP in (\ref{eq:LinearProgram}) to obtain $\{\lp{\li}{\lj}\}$. Then, Algorithm \ref{A1} creates a set, $\ls$, of slots (see Section~\ref{subsec:SA1SlotAlloc}). After we have the set $\ls$, Algorithm \ref{A2} computes the service placement $\rpt{\fsa}$. In Algorithm \ref{A2}, the objects $\psa$ and $\fsa$ are maps from $\ls$ into $\mathbb{N}\cup\{\emptyset\}$. The algorithm has a function $\exr{\cdot}$ which takes, as input, a map from $\ls$ into $\mathbb{N}\cup\{\emptyset\}$. This function $\exr{\cdot}$ is computed in Algorithm \ref{A3}.

In the rest of this section, we give a description of the mechanics of the algorithm and a proof of the approximation ratio and feasibility of the computed service placement $\rpt{\fsa}$.

\subsection{Slots Allocations}
\label{subsec:SA1SlotAlloc}

A \textit{slot} $\su$ is an object that has two associated values: $\sv{\su}\in\nd$ and $\sw{\su}\in\mathbb{N}$. Intuitively, a slot $\su$ is a space, with capacity $\fc^{\sw{\su}-1}\ca{\lj}\fb$, on node $\sv{\su}$. A slot $\su$ will hold a single service in $\pa{\sv{\su}}{\sw{\su}}$. For all $\lj\in\nd$ and $\ld\in\mathbb{N}$, Algorithm \ref{A1} creates $\ns{\lj}{\ld}$ slots $\su$ with $\sv{\su}:=\lj$ and $\sw{\su}:=\ld$. $\ls$ is the set of all slots created.

A \textit{slot allocation} $\sla$ is a function from $\ls$ into $\sr$ such that, given a slot $\su\in\ls$, we have $\sla(\su)\in\pa{\sv{\su}}{\sw{\su}}$. A slot allocation $\sla$ is an assignment of services to slots such that given a slot $\su$, the service $\sla(\su)$ assigned to it is in $\pa{\sv{\su}}{\sw{\su}}$, implying that $\sla(\su)$ does not exceed the capacity of $\su$.

A \textit{partial slot allocation} $\psa$ is a function from $\ls$ into $\sr\cup\{\emptyset\}$ such that, given a slot $\su\in\ls$, we have $\psa(\su)\in\pa{\sv{\su}}{\sw{\su}}\cup\{\emptyset\}$. A partial slot allocation $\psa$ is a partial assignment of services to slots: given a slot $\su$, $\psa(\su)=\emptyset$ means that no service has been assigned to $\su$ (i.e. the slot is empty), and $\psa(\su)\neq\emptyset$ means that service $\psa(\su)$ has been assigned to $\su$ (and, as for slot allocations, we have $\psa(\su)\in\pa{\sv{\su}}{\sw{\su}}$).

Given a partial slot allocation $\psa$ we define $\sfa{\psa}$ as the set of all slot allocations $\sla$, where $\sla(\su)=\psa(\su)$ for all $\su\in\ls$ with $\psa(\su)\neq\emptyset$.  $\sfa{\psa}$ is the set of all slot allocations that can be obtained by assigning services to all the empty slots of $\psa$.

Given any slot allocation $\sla$ we define its \textit{associated service placement}, $\rpt{\sla}$, by: $$\set{\sla}{\li}:=\{\lj:\exists~\su\in\ls\operatorname{~with~}\sv{\su}=\lj\operatorname{~and~}\sla(\su)=\li\}$$
for all $\li\in\sr$. This means that service $\li$ is placed on node $\lj$ iff there exists a slot on node $\lj$ which contains service $\li$.

Given a slot allocation $\sla$ and a user $\lk\in\ur$, we define $\sat{\sla}{\lk}:=1$ if there exists a node $\lj$ in $\us{\lk}$ and a slot $\su\in\ls$ with $\sv{\su}=\lj$ and $\sla(\su)=\ui{\lk}$, and $\sat{\sla}{\lk}:=0$ otherwise. Note that:
\begin{equation}
\sat{\sla}{\lk}=\id{\us{\lk}\cap\set{\sla}{\ui{\lk}}\neq\emptyset}. \label{eq:def_f}
\end{equation}

\begin{theorem}\label{FSA}
For any slot allocation $\sla$ (with set of slots $\ls$), its associated service placement $\rpt{\sla}$ is feasible.
\end{theorem}

\begin{prf}
        First note that for all $\lj\in\nd$ and $\li\in\sr$ we have: $$\id{\lj\in\set{\sla}{\li}}=\id{\exists~\su\in\ls\operatorname{~with~}\sv{\su}=\lj\operatorname{~and~}\sla(\su)=\li}$$
        so for all $\lj\in\nd$ we have:
        {\footnotesize
        \begin{align*}
        &\sum_{\li\in\sr}\sz{\li}\cdot\id{\lj\in\set{\sla}{\li}} 
        \leq~\sum_{\li\in\sr}~\sum_{\su\in\ls:\sv{\su}=\lj,\sla(\su)=\li}\sz{\li}
        =\sum_{\su\in\ls:\sv{\su}=\lj}\sz{\sla(\su)} \\
        &=\sum_{\ld\in\mathbb{N}}~\sum_{\su\in\ls:\sv{\su}=\lj, \sw{\su}=\ld}\sz{\sla(\su)} 
        \stackrel{\textrm{\textcircled{a}}}{\leq} \sum_{\ld\in\mathbb{N}}~\sum_{\su\in\ls:\sv{\su}=\lj, \sw{\su}=\ld}\fc^{\ld-1}\ca{\lj}\fb  \\
        &\stackrel{\textrm{\textcircled{b}}}{\leq} \sum_{\ld\in\mathbb{N}}\left(\vo{\lj}\sum_{\li\in\pa{\lj}{\ld}}\lp{\li}{\lj}+1\right)\fc^{\ld-1}\ca{\lj}\fb \\
        &=\ca{\lj}\fb\left(\sum_{\ld\in\mathbb{N}}\fc^{\ld-1}\right)+\frac{\vo{\lj}}{\fc}\left(\sum_{\ld\in\mathbb{N}}\sum_{\li\in\pa{\lj}{\ld}}\lp{\li}{\lj}\fc^{\ld}\fb\ca{\lj}\right)\\
        &=\frac{\ca{\lj}\fb}{1-\fc}\!+\!\frac{\vo{\lj}}{\fc}\left(\sum_{\ld\in\mathbb{N}}\sum_{\li\in\pa{\lj}{\ld}}\!\!\lp{\li}{\lj}\fc^{\ld}\fb\ca{\lj}\!\!\right)
        \leq\frac{\ca{\lj}\fb}{1-\fc}\!+\!\frac{\vo{\lj}}{\fc}\left(\sum_{\ld\in\mathbb{N}}\sum_{\li\in\pa{\lj}{\ld}}\!\!\lp{\li}{\lj}\sz{\li}\!\!\right)\\
        &=\frac{\ca{\lj}\fb}{1-\fc}+\frac{\vo{\lj}}{\fc}\left(\sum_{\li\in\sr}\lp{\li}{\lj}\sz{\li}\right)
        \leq\frac{\ca{\lj}\fb}{1-\fc}+\frac{\fd\ca{\lj}}{\fc}
        =\ca{\lj}
        \end{align*}
        }%
        which proves the feasibility of $\rpt{\sla}$.
        In the above, step \textcircled{a} is because for all $\su\in\ls$ with $\sv{\su}=\lj$ and $\sw{\su}=\ld$, we have $\sla(\su)\in\pa{\lj}{\ld}$; step \textcircled{b} is because  $|\{\su\in\ls:\sv{\su}=\lj,\sw{\su}=\ld\}|=\ns{\lj}{\ld}\leq\vo{\lj}\sum_{\li\in\pa{\lj}{\ld}}\lp{\li}{\lj}+1$.
        The other steps are mainly from the definitions in (\ref{eq:defGammaDelta}) and Definition~\ref{def:SA1}.
\end{prf}

\subsection{Probability Distribution for SA1}
\label{subsec:ProbDistributionSA1}
For the analysis of SA1, we define a probability distribution on the set of possible slot allocations. This will guide, via the method of conditional expectations~\cite{vazirani2013approximation}, the construction of the slot allocation $\fsa$ in Algorithm \ref{A2}.
The probability distribution on slot allocations $\sla$ is defined as follows: for every slot $\su\in\ls$ independently, draw a service $\li$ from $\pa{\sv{\su}}{\sw{\su}}$ with probability $\lp{\li}{\sv{\su}}/\de{\sv{\su}}{\sw{\su}}$ and set $\sla(\su)\leftarrow\li$.

\begin{definition}
\label{def:CondExpectSA1}
Given a partial slot allocation $\psa$, define: 
$$\exr{\psa}:=\mathbb{E}\left(\sum_{\lk\in\ur}\sat{\sla}{\lk}\uw{\lk}\Bigg|\sla\in\sfa{\psa}\right)$$
where $\mathbb{E}$ denotes the expectation.
$\exr{\cdot}$ appears in Algorithm~\ref{A2} and is computed in Algorithm~\ref{A3} (see Theorem~\ref{theorem:SA1ExpectCorrectness}).
\end{definition}

The next theorem shows that when we draw $\sla$ from the above probability distribution, the expected total reward of its associated service placement is bounded below by $(1-e^{-\fd})\rw$.

\begin{theorem}\label{ER}
Under our probability distribution for SA1, the expected total reward $\mathbb{E}\left(\sum_{\lk\in\ur}\sat{\sla}{\lk}\uw{\lk}\right) \geq (1-e^{-\fd})\rw$.
\end{theorem}

\begin{prf}
    First note that:
    \footnotesize
    $$\mathbb{E}\left(\sum_{\lk\in\ur}\sat{\sla}{\lk}\uw{\lk}\right)=\sum_{\lk\in\ur}\uw{\lk}\mathbb{E}(\sat{\sla}{\lk}) = \sum_{\lk\in\ur}\uw{\lk}\pr{\sat{\sla}{\lk}=1}$$ 
    \normalsize
    where the last equality is since $\sat{\sla}{\lk}$ is boolean.  
    We shall now bound $\pr{\sat{\sla}{\lk}=1}$. Note first that $\sat{\sla}{\lk}=1$ if there exists a slot $\su\in\ls$ with $\sv{\su}\in\us{\lk}$ and $\sla(\su)=\ui{\lk}$ so, since $\sla(\su)$ is drawn independently for every slot $\su$, we have, by%
    \if\citetechreport1
        Lemma~\ref{p1} in~\cite[Appendix~\ref{append:Lemmas}]{techReportThisPaper},
    \else
        ~Lemma~\ref{p1} in Appendix~\ref{append:Lemmas},
    \fi
    that:
    
    \vspace{-0.15in}
    \footnotesize 
    $$\pr{\sat{\sla}{\lk}=1}\geq\left(1-\exp\left(-\sum_{\su\in\ls:\sv{\su}\in\us{\lk}}\pr{\sla(\su)=\ui{\lk}}\right)\right).$$
    \normalsize
    Given a node $\lj\in\nd$ define $\slb{\lj}{\lk}$ such that $\ui{\lk}\in\pa{\lj}{(\slb{\lj}{\lk})}$. We then have
    \footnotesize
    \begin{align*}
    &\sum_{\su\in\ls:\sv{\su}\in\us{\lk}}\pr{\sla(\su)=\ui{\lk}}
    = \sum_{\lj\in\us{\lk}}~\sum_{\su\in\ls:\sv{\su}=\lj}\pr{\sla(\su)=\ui{\lk}}\\
    &=\sum_{\lj\in\us{\lk}}~\sum_{\su\in\ls:\sv{\su}=\lj, \sw{\su}=\slb{\lj}{\lk}}\pr{\sla(\su)=\ui{\lk}}\\
    &=\sum_{\lj\in\us{\lk}}~\sum_{\su\in\ls:\sv{\su}=\lj, \sw{\su}=\slb{\lj}{\lk}}\lp{\ui{\lk}}{\sv{\su}}/\de{\sv{\su}}{\sw{\su}}\\
    &=\sum_{\lj\in\us{\lk}}~\sum_{\su\in\ls:\sv{\su}=\lj, \sw{\su}=\slb{\lj}{\lk}}\lp{\ui{\lk}}{\lj}/\de{\lj}{(\slb{\lj}{\lk})}\\
    &=\sum_{\lj\in\us{\lk}}~\frac{\lp{\ui{\lk}}{\lj}}{\de{\lj}{(\slb{\lj}{\lk})}}\sum_{\su\in\ls:\sv{\su}=\lj, \sw{\su}=\slb{\lj}{\lk}}1\\
    &=\sum_{\lj\in\us{\lk}}~\frac{\lp{\ui{\lk}}{\lj}}{\de{\lj}{(\slb{\lj}{\lk})}}\ns{\lj}{(\slb{\lj}{\lk})}\\
    &=\sum_{\lj\in\us{\lk}}~\frac{\lp{\ui{\lk}}{\lj}}{\de{\lj}{(\slb{\lj}{\lk})}}\left\lceil\fd\de{\lj}{(\slb{\lj}{\lk})}\right\rceil
    \geq \fd\sum_{\lj\in\us{\lk}}~\lp{\ui{\lk}}{\lj}
    \geq \fd\lr{\lk}.
    \end{align*}
    \normalsize
    Plugging into the above, we have:
    $$\pr{\sat{\sla}{\lk}=1}\geq 1-\exp(-\fd\lr{\lk}).$$
    Since $\lr{\lk}\leq 1$,%
    \if\citetechreport1
        by Lemma~\ref{p2} in~\cite[Appendix~\ref{append:Lemmas}]{techReportThisPaper},
    \else
        ~by Lemma~\ref{p2} in Appendix~\ref{append:Lemmas},
    \fi
    we have $1-\exp(-\fd\lr{\lk}) \geq \lr{\lk}(1-e^{-\fd})$ which gives us an expected total reward of: 
    \footnotesize
    $$\sum_{\lk\in\ur}\uw{\lk}\lr{\lk}(1-e^{-\fd})=(1-e^{-\fd})\arw\geq (1-e^{-\fd})\rw$$
    \normalsize
    where the last equality is from Theorem \ref{BOR}.
\end{prf}

\emph{Remark:} The randomized step is only needed for theoretical analysis. It \emph{does not} exist in the algorithm. Our algorithm (See Algorithms~\ref{A1}, \ref{A2}, and \ref{A3}) only needs to compute the expected value given in Definition~\ref{def:CondExpectSA1} and \emph{does not} include any randomized step. Hence, \emph{our algorithm SA1 is deterministic}. The same applies to SA2 and other algorithms presented later.

\subsection{Placing Services}\label{AL1}
We now describe and analyze Algorithm \ref{A2}, which uses the idea of the method of conditional expectations~\cite{vazirani2013approximation}. 

In Algorithm \ref{A2}, we maintain a partial slot allocation $\psa$ where, initially, all slots are empty. Every time we go around the loop in Lines \ref{lia}-\ref{lib} of Algorithm \ref{A2}, we do the following:
\begin{enumerate}
\item Pick an empty slot $\su$.
\item For all $\li\in\pa{\sv{\su}}{\sw\su}$ define $\ex{\sla}{\li}$ to be the partial slot allocation formed from $\psa$ by assigning the service $\li$ to the slot $\su$.
\item Choose $\li'\in\pa{\sv{\su}}{\sw\su}$ that maximizes $\exr{\ex{\sla}{\li'}}$.
\item Update $\psa$ by assigning service $\li'$ to the slot $\su$.
\end{enumerate}
We loop through the above until all slots have been assigned services (i.e., $\psa$ is a slot allocation). We let $\fsa$ be the slot allocation that has now been constructed and output its associated service placement $\rpt{\fsa}$.
Theorem \ref{FSA} shows that $\rpt{\fsa}$ is feasible. We now prove the approximation ratio.

\begin{theorem}\label{EXR1}
The service placement $\rpt{\fsa}$, computed by Algorithm \ref{A2} with slot creation computed by Algorithm~\ref{A1}, has a total reward of at least $\mathbb{E}\left(\sum_{\lk\in\ur}\sat{\sla}{\lk}\uw{\lk}\right)$.
\end{theorem}

\begin{prf}
We maintain the inductive hypothesis that $\exr{\psa}\geq\mathbb{E}\left(\sum_{\lk\in\ur}\sat{\sla}{\lk}\uw{\lk}\right)$ throughout the algorithm. 

Initially we have $\psa(\su)=\emptyset$ for every $\su\in\ls$, thus $\sfa{\psa}$ is the set of all possible slot allocations and hence $\exr{\psa}=\mathbb{E}\left(\sum_{\lk\in\ur}\sat{\sla}{\lk}\uw{\lk}\right)$, so the inductive hypothesis holds.

Now suppose that the inductive hypothesis holds at some point during the algorithm. After Line~\ref{alg:ServicePlacement:computeMaxExpect} and before Line~\ref{lib} of Algorithm~\ref{A2}, we have:

\vspace{-0.15in}
{\footnotesize
\begin{align*}
\exr{\ex{\sla}{\li'}}&=\!\!\max_{\li\in\pa{\sv{\su}}{\sw{\su}}}\!\!\exr{\ex{\sla}{\li}} 
 =\!\!\max_{\li\in\pa{\sv{\su}}{\sw{\su}}}\!\! \mathbb{E}\!\left(\sum_{\lk\in\ur}\sat{\sla}{\lk}\uw{\lk}\Bigg|\sla\in\sfa{\ex{\sla}{\li}}\!\right)\\
& \geq~\mathbb{E}\left(\sum_{\lk\in\ur}\sat{\sla}{\lk}\uw{\lk}\Bigg|\sla\in\sfa{\psa}\right)
  =\exr{\psa}
 \end{align*} }%
 where the inequality is because  $\{\sfa{\ex{\sla}{\li}}:\li\in\pa{\sv{\su}}{\sw{\su}}\}$ is a partition of $\sfa{\psa}$.
By the inductive hypothesis, this is bounded below by $\mathbb{E}\left(\sum_{\lk\in\ur}\sat{\sla}{\lk}\uw{\lk}\right)$. The fact that $\psa$ is then updated by $\ex{\sla}{\li'}$ in Line~\ref{lib} of Algorithm~\ref{A2} proves the inductive hypothesis.

The inductive hypothesis hence holds always which means, as $\sfa{\fsa}=\{\fsa\}$, we have $\sum_{\lk\in\ur}\sat{\fsa}{\lk}\uw{\lk}=\exr{\fsa}\geq\mathbb{E}\left(\sum_{\lk\in\ur}\sat{\sla}{\lk}\uw{\lk}\right)$, which proves the result.
\end{prf}

\begin{theorem} \label{theorem:SA1ApproxRatio}
The service placement $\rpt{\fsa}$, computed by Algorithm \ref{A2} with slot creation computed by Algorithm~\ref{A1}, has a total reward of at least $(1-e^{-\fd})\rw$.
\end{theorem}

\begin{prf}
The result is direct from Theorems \ref{ER} and \ref{EXR1}
\end{prf}

We now prove the correctness of Algorithm \ref{A3}.
\begin{theorem} \label{theorem:SA1ExpectCorrectness}
Algorithm \ref{A3} computes $\exr{\ex{\sla}{}}$ correctly.
\end{theorem}

\begin{prf}
Algorithm \ref{A3} first computes the value: $$\pf{\lk}:=\pr{\sat{\sla}{\lk}=1|\sla\in\sfa{\ex{\sla}{}}}.$$
The fact that the algorithm computes the correct value of $\pf{\lk}$ can be seen as follows:

If there exists a node $\lj\in\us{\lk}$ and a slot $\su$ with $\sv{\su}=\lj$ and $\ex{\sla}{}(\su)=\ui{\lk}$, then by definition of $\sfa{\ex{\sla}{}}$, every service placement $\sla$ in $\sfa{\ex{\sla}{}}$ has $\sla(\su)=\ui{\lk}$. Therefore, by definition of $\sat{\sla}{\lk}$, we have $\sat{\sla}{\lk}=1$ for all $\sla$ in $\sfa{\ex{\sla}{}}$. This implies that $\pf{\lk}:=\pr{\sat{\sla}{\lk}=1|\sla\in\sfa{\ex{\sla}{}}}=1$.

We now consider the situation where there does not exist a node $\lj\in\us{\lk}$ and a slot $\su$ with $\sv{\su}=\lj$ and $\ex{\sla}{}(\su)=\ui{\lk}$. For all $\sla\in\sfa{\ex{\sla}{}}$, we have $\sat{\sla}{\lk}=1$ iff there exists some $\lj\in\us{\lk}$ and slot $\su$ with $\sla(\su)=\ui{\lk}$. Thus, since each slot is filled independently (even under the condition $\sla\in\sfa{\ex{\sla}{}}$) we have, by Lemma \ref{p5}%
\if\citetechreport1
in~\cite[Appendix~\ref{append:Lemmas}]{techReportThisPaper},
\else
~in Appendix~\ref{append:Lemmas},
\fi
that:

\vspace{-0.1in}
{\footnotesize
\begin{align*}
\pf{\lk}&=\pr{\sat{\sla}{\lk}=1|\sla\in\sfa{\ex{\sla}{}}}
=1-\!\!\!\!\!\!\!\! \prod_{\su\in\ls:\sv{\su}\in\us{\lk}}\!\!\!\!\!\!(1\!-\!\pr{\sla(\su)=\ui{\lk}|\sla\in\sfa{\ex{\sla}{}}}).
\end{align*}  }%
For any slot $\su\in\ls$ with $\psa(\su)\neq\emptyset$ or $\ui{\lk}\notin\pa{\lj}{\sw{\su}}$, we have $1-\pr{\sla(\su)=\ui{\lk}|\sla\in\sfa{\ex{\sla}{}}}=1$, thus we can remove it from the product above. Noting then that for all other $\su$ we have $1-\pr{\sla(\su)=\ui{\lk}|\sla\in\sfa{\ex{\sla}{}}}=1-\lp{\li}{\lj}/\de{\lj}{\sw{\su}}$, we have shown the correctness of the computation of $\pf{\lk}$ by Algorithm \ref{A3}.

We now have:

\vspace{-0.13in}
{\footnotesize
\begin{align*}
\exr{\ex{\sla}{}}&=\mathbb{E}\left(\sum_{\lk\in\ur}\sat{\sla}{\lk}\uw{\lk}\Bigg|\sla\in\sfa{\ex{\sla}{}}\right) =\sum_{\lk\in\ur}\uw{\lk}\mathbb{E}(\sat{\sla}{\lk}|\sla\in\sfa{\ex{\sla}{}})\\
&=\sum_{\lk\in\ur}\uw{\lk}\pr{\sat{\sla}{\lk}=1 \big|\sla\in\sfa{\ex{\sla}{}}} =\sum_{\lk\in\ur}\uw{\lk}\pf{\lk}
\end{align*}  }%
which proves the correctness of Algorithm \ref{A3}.
\end{prf}

\section{Second Slot Allocation Algorithm (SA2)}
\label{sec:SA2}

The second algorithm, SA2, for solving SPSC is for the case where services can be arbitrarily large. We fix $\fb$ equal to 1/4.
From (\ref{eq:defGammaDelta}), we immediately get $\fc = 1/2$ and $\fd = 1/4$.
SA2 has an approximation ratio of $\left(1-e^{-1}\right) / 4$, which will be given by Theorem~\ref{theorem:SA2ApproxRatio} later.

\begin{definition}
After solving the LP in (\ref{eq:LinearProgram})we define the following for SA2, for all $\lj\in\nd$:

\vspace{-0.13in}
{\footnotesize
\begin{align*}
\pa{\lj}{\bsl}&:=\{\li\in\sr:\frac{1}{2}\ca{\lj}<\sz{\li}\leq\ca{\lj}\};  \\
\pa{\lj}{\ssl}&:=\{\li\in\sr:\ca{\lj}\fb<\sz{\li}\leq\frac{1}{2}\ca{\lj}\};  \\
 \pa{\lj}{\ld}&:=\{\li\in\sr:\fc^{\ld}\ca{\lj}\fb<\sz{\li}\leq\fc^{\ld-1}\ca{\lj}\fb\} ,&&  \forall \ld\in\mathbb{N}; \\
\de{\lj}{\ld}&:=\sum_{\li\in\pa{\lj}{\ld}}\lp{\li}{\lj} , &&\forall \ld\in\mathbb{N}\cup\{\bsl,\ssl\}; \\
\vo{\lj}&:=\frac{\fd\ca{\lj}}{\sum_{\li\in\sr:\sz{\li}\leq\ca{\lj}\fb}\sz{\li}\lp{\li}{\lj}};  \\
\ns{\lj}{\ld}&:=\left\lceil\vo{\lj}\de{\lj}{\ld}\right\rceil ,&& \forall \ld\in\mathbb{N}.
\end{align*}
}%

We also define, for all $\lj\in\nd$, the quantity $\pq{\lj}$ as follows:
\begin{itemize}
\item If $\de{\lj}{\ssl}<2$, then $\pq{\lj}:=\de{\lj}{\ssl}$.
\item  If $\de{\lj}{\ssl}\geq 2$, then $\pq{\lj}:=\de{\lj}{\ssl}/2$. 
\end{itemize}
\end{definition}

The only difference between SA1 and SA2 is in the slots created by the respective slot creation algorithms. In SA2, a slot $\su$ is again an object with two associated values: $\sv{\su}$ and $\sw{\su}$, except that in SA2, we now have $\sw{\su}\in\mathbb{N}\cup\{\bsl,\ssl\}$.

\textbf{Procedure of SA2:} First, we solve the LP in (\ref{eq:LinearProgram}) to obtain $\{\lp{\li}{\lj}\}$. Then, Algorithm \ref{A4} creates a set, $\ls$, of slots. After we have the set $\ls$, Algorithm \ref{A2} (in Section~\ref{sec:SA1}) computes the service placement $\rpt{\fsa}$. In Algorithm \ref{A4}, the objects $\pap$ and $\fm$ are maps from $\nd$ into $\{\ha,\hb,\hc,\emptyset\}$. The algorithm has a function $\exs{\cdot}$ which takes, as input, a map from $\nd$ into $\{\ha,\hb,\hc,\emptyset\}$. This function $\exs{\cdot}$ is computed in Algorithm \ref{A5}.

Next, we give a description of the mechanics of the algorithm and present theoretical results on the approximation ratio and feasibility of the computed service placement $\rpt{\fsa}$.

\subsection{Construction Maps}\label{CM}
A \textit{construction map}, $\map$, is a map from $\nd$ to $\{\ha,\hb,\hc\}$. Intuitively, a construction map $\map$ is a labelling of all nodes by the label $\ha$, $\hb$ or $\hc$.

A \textit{partial construction map}, $\pap$, is a map from $\nd$ to $\{\ha,\hb,\hc,\emptyset\}$. Intuitively a partial construction map $\pap$ is a partial labelling of the nodes by labels $\ha$, $\hb$, $\hc$. If, for node $\lj\in\nd$, $\pap(\lj)=\emptyset$, then $\lj$ has no label. Otherwise,  $\lj$ has label $\pap(\lj)$.

Given a partial construction map $\pap$, we define $\mas{\pap}$ to be the set of all construction maps $\map$ such that $\map(\lj)=\pap(\lj)$ for all $\lj\in\nd$ with $\pap(\lj)\neq\emptyset$. Intuitively, $\mas{\pap}$ is the set of all construction maps that can be obtained from $\pap$ by assigning labels to the unlabelled nodes.

Every construction map has an associated set of slots defined as follows, for all $\lj\in\nd$:
\begin{itemize}
\item If $\map(\lj)=\ha$, we have a single slot $\su$ with $\sv{\su}:=\lj$ and $\sw{\su}:=\bsl$. There are no other slots $\su'$ with $\sv{\su'}=\lj$.
\item If $\map(\lj)=\hb$, we have two slots $\su$ with $\sv{\su}:=\lj$ and $\sw{\su}:=\ssl$. There are no other slots $\su'$ with $\sv{\su'}=\lj$.
\item If $\map(\lj)=\hc$, then for all $\ld\in\mathbb{N}$, we have $\ns{\lj}{\ld}$ slots $\su$ with $\sv{\su}:=\lj$ and $\sw{\su}:=\ld$. There are no other slots $\su'$ with $\sv{\su'}=\lj$.
\end{itemize}

\begin{algorithm}
\caption{Slot Creation of SA2}
\label{A4}
\begin{algorithmic}[1]
  \footnotesize
  \STATE For every $\lj\in\nd$: set $\pap(\lj)\leftarrow\emptyset$;
  \STATE For every $\lj\in\nd$:
  \STATE \label{2cc}~~~~~~~  For all $a\in\{\ha,\hb,\hc\}$:
  \STATE ~~~~~~~~~~~~~~ $\ex{\map}{a}(\lj)\leftarrow a$;
  \STATE ~~~~~~~~~~~~~~ For all $\lj'\in\nd\setminus\{\lj\}$: set $\ex{\map}{a}(\lj')\leftarrow\pap(\lj')$;
  \STATE ~~~~~~~ $a'\leftarrow\operatorname{argmax}_{a\in\{\ha,\hb,\hc\}}\exs{\ex{\map}{a}}$;
  \STATE\label{2cd} ~~~~~~~ $\pap(\lj)\leftarrow a'$ ;
  \STATE $\fm\leftarrow\pap$;
  \STATE \label{2ce}For all $\lj\in\nd$:
  \STATE ~~~~~~~ If $\fm(\lj)=\ha$: create a single slot $\su$ with $\sv{\su}\leftarrow\lj$, $\sw{\su}\leftarrow\bsl$;
  \STATE ~~~~~~~ If $\fm(\lj)=\hb$: create two slots $\su$ with $\sv{\su}\leftarrow\lj$, $\sw{\su}\leftarrow\ssl$;
  \STATE~~~~~~~  If $\fm(\lj)=\hc$ then: for all $\ld\in\mathbb{N}$ with $\pa{\lj}{\ld}\neq\emptyset$: 
  \STATE \label{2cf}~~~~~~~~~~~~~~Create $\ns{\lj}{\ld}$ slots $\su$ with $\sv{\su}\leftarrow\lj$ and $\sw{\su}\leftarrow\ld$;
  \STATE Output $\ls$ as the set of all slots created;
\end{algorithmic}
\end{algorithm}

\begin{algorithm}
\caption{Computing $\exs{\ex{\map}{}}$}
\label{A5}
\begin{algorithmic}[1]
  \footnotesize
  \STATE For all $\li\in\sr$, $\lj\in\nd$:
  \STATE ~~~~~~~If $\sz{\li}>\ca{\lj}$ then: $\pbl{\li}{\lj}{a}\leftarrow 0$ for all $a\in\{\ha,\hb,\hc\}$;
  \STATE~~~~~~~If $\li\in\pa{\lj}{\bsl}$ then:
  \STATE~~~~~~~~~~~~~~ $\pbl{\li}{\lj}{\ha}\leftarrow\lp{\li}{\lj}/\de{\lj}{\bsl}$;
   \STATE~~~~~~~~~~~~~~  For all $a\in\{\hb,\hc\}$: set $\pbl{\li}{\lj}{a}\leftarrow 0$;
  \STATE~~~~~~~If $\li\in\pa{\lj}{\ssl}$ then:
    \STATE~~~~~~~~~~~~~~ $\pbl{\li}{\lj}{\hb}\leftarrow(1-(1-\lp{\li}{\lj}/\de{\lj}{\ssl})^2)$;
   \STATE~~~~~~~~~~~~~~  For all $a\in\{\ha,\hc\}$: set $\pbl{\li}{\lj}{a}\leftarrow 0$;
  \STATE~~~~~~~If $\exists\ld\in\mathbb{N}$ with $\li\in\pa{\lj}{\ld}$ then:
   \STATE~~~~~~~~~~~~~~ $\pbl{\li}{\lj}{\hc}\leftarrow(1-(1-\lp{\li}{\lj}/\de{\lj}{\ld})^{\ns{\lj}{\ld}})$;
   \STATE~~~~~~~~~~~~~~ For all $a\in\{\ha,\hb\}$: set $\pbl{\li}{\lj}{a}\leftarrow 0$;
   \STATE~~~~~~~$\pbl{\li}{\lj}{\emptyset}\leftarrow\fd\de{\lj}{\bsl}\pbl{\li}{\lj}{\ha}+\fd\pq{\lj}\pbl{\li}{\lj}{\hb}+(1-\fd\de{\lj}{\bsl}-\fd\pq{\lj})\pbl{\li}{\lj}{\hc}$;
   \STATE For all $\lk\in\ur$: 
   \STATE~~~~~~~$p\leftarrow 1$;
  \STATE~~~~~~~For all $\lj\in\us{\lk}$:
   \STATE~~~~~~~~~~~~~~$p\leftarrow p(1-\pbl{\ui{\lk}}{\lj}{\ex{\map}{}(\lj)})$;
  \STATE~~~~~~~$\pg{\lk}=1-p$;
  \STATE Output $\exs{\ex{\map}{}}\leftarrow\sum_{\lk\in\ur}\pg{\lk}\uw{\lk}$;
\end{algorithmic}
\end{algorithm}

\subsection{Probability Distribution for SA2}\label{PD}
For the theoretical analysis of SA2, we define a probability distribution over the set of slots $\ls$ as well as the slot allocation. Similar to the analysis of SA1, this probability distribution will guide the construction of the set, $\ls$, of slots in Algorithm \ref{A4}.

We first define a probability distribution over a map $\map:\nd\rightarrow\{\ha,\hb,\hc\}$ as follows: for each $\lj\in\nd$ independently, set $\map(\lj)\leftarrow \ha$ with probability $\fd\de{\lj}{\bsl}$, set $\map(\lj)\leftarrow \hb$ with probability $\fd\pq{\lj}$, and set $\map(\lj)\leftarrow \hc$ with probability $1-\fd\de{\lj}{\bsl}-\fd\pq{\lj}$.
Once $\map$ has been sampled from this probability distribution, we let $\ls$ be its associated set of slots (defined in Section \ref{CM}).
For the selected $\ls$, we define the probability distribution over slot allocations, $\sla$, in the same way as for SA1 in Section~\ref{subsec:ProbDistributionSA1}. 

\begin{definition}
Define: 

\vspace{-0.13in}
{\footnotesize
$$\exs{\pap}:=\mathbb{E}\left(\sum_{\lk\in\ur}\sat{\sla}{\lk}\uw{\lk}\Bigg|\map\in\mas{\pap}\right)$$ }%
where $\sat{\sla}{\lk}$ is defined as in (\ref{eq:def_f}). $\exs{\cdot}$ appears in Algorithm \ref{A4} and is computed in Algorithm \ref{A5} (see Theorem~\ref{theorem:CorrectnessOfExpectationSA2}).
\end{definition}

\if\citetechreport1
The proofs of some of the theorems presented next are included in our online technical report~\cite{techReportThisPaper}.
\fi

\begin{theorem}
\label{theorem:FeasibilitySA2}
Under our probability distribution for SA2, for any slot allocation $\sla$ of non-zero probability, its associated service placement, $\rpt{\sla}$, is feasible. 
\end{theorem}
\if\citetechreport0
\begin{prf}
    See Appendix~\ref{append:ProofFeasibleSA2}.
\end{prf}
\fi

\begin{theorem}\label{ER2}
Under our probability distribution for SA2, the expected total reward $\mathbb{E}\left(\sum_{\lk\in\ur}\sat{\sla}{\lk}\uw{\lk}\right) \geq \left(1-e^{-1}\right)\fd\rw$.
\end{theorem}

\if\citetechreport0
\begin{prf}
    See Appendix~\ref{append:ProofExpectedLowerBoundSA2}.
\end{prf}
\fi

\subsection{Creating Slots}

In Algorithm~\ref{A4}, we maintain a partial construction map $\pap$ where, initially, all nodes are unlabelled. Every time we run the loop in Lines \ref{2cc}-\ref{2cd} of Algorithm~\ref{A4}, we do the following:
\begin{enumerate}
\item Pick an unlabelled node $\lj$.
\item For all $a\in\{1,2,3\}$, define $\ex{\map}{a}$ as the partial construction map formed from $\pap$ by labelling $\lj$ with $a$.
\item Choose $a'\in\{\ha,\hb,\hc\}$ that maximizes $\exs{\ex{\map}{a'}}$.
\item Update $\pap$ by labelling $\lj$ with $a'$.
\end{enumerate}
We loop through the above until all nodes have been labelled. Then, let $\fm$ be the construction map we have made.  Lines \ref{2ce}-\ref{2cf} of Algorithm~\ref{A4} then create its associated set of slots, $\ls$.

\begin{theorem}\label{ET}
$\mathbb{E}\left(\sum_{\lk\in\ur}\sat{\sla}{\lk}\uw{\lk}|\map \!= \!\fm\right)\geq\mathbb{E}\left(\sum_{\lk\in\ur}\sat{\sla}{\lk}\uw{\lk}\right)$.
\end{theorem}

\if\citetechreport0
\begin{prf}
    See Appendix~\ref{append:ProofMethodOfCondExpectSA2}.
\end{prf}
\fi

\begin{theorem}\label{EXR2}
The service placement $\rpt{\fsa}$, computed by Algorithm \ref{A2} with slot creation computed by Algorithm~\ref{A4}, has a total reward of at least $\mathbb{E}\left(\sum_{\lk\in\ur}\sat{\sla}{\lk}\uw{\lk}|\map=\fm\right)$.
\end{theorem}

\begin{prf}
As in the proof of Theorem \ref{EXR1}, noting that the service placement is computed, from $\ls$, by Algorithm \ref{A2}.
\end{prf}

\begin{theorem} \label{theorem:SA2ApproxRatio}
The service placement $\rpt{\fsa}$, computed by Algorithm \ref{A2} with slot creation computed by Algorithm~\ref{A4}, has a total reward of at least $(1-e^{-1})\fd\rw$.
\end{theorem}

\begin{prf}
Directly from Theorems \ref{ER2}, \ref{ET}, and \ref{EXR2}.
\end{prf}

\begin{theorem}
\label{theorem:CorrectnessOfExpectationSA2}
Algorithm \ref{A5} computes $\exs{\ex{\map}{}}$ correctly.
\end{theorem}
\if\citetechreport0
\begin{prf}
    See Appendix~\ref{append:ProofCorrectnessOfExpectationSA2}.
\end{prf}
\fi

\section{Overall Algorithm}
\label{sec:OverallAlgorithm}

\subsection{Combining the Algorithms and Repeating}
We now present the overall algorithm for SPSC, which has better empirical performance than SA1 and SA2 alone. First, we note that SA1 and SA2 can be combined as follows. Find the minimum $\beta$ such that $\max_{\li\in\sr}\sz{\li}\leq\fb (\min_{\lj\in\nd}\ca{\lj})$. If $\beta \geq 1$ or $\left(1-e^{-1}\right) / 4 > 1-e^{-\left(1-\sqrt{\fb}\right)^2}$, run SA2; else, run SA1. We call this the combined slot allocation algorithm (CSA). 

We found empirically that after running CSA, many nodes have an excessive amount of capacity remaining, i.e., $\sum_{\li:\lj\in\se{\li}}\sz{\li}$ is significantly smaller than $\ca{\lj}$. This is because SA1 and SA2, thus CSA, both guarantee feasibility and may under-utilize the nodes. We now give a repeated slot allocation algorithm (RSA) which utilizes the remaining capacity by repeating CSA, as shown in Algorithm~\ref{alg:RSA}.

\begin{algorithm}
\caption{Repeated slot allocation (RSA)}
\label{alg:RSA}
\begin{algorithmic}[1]
  \footnotesize
  \STATE For all $\li\in\sr$: set $Y_{\li}\leftarrow\emptyset$;
  \STATE Set $\ur'\leftarrow\ur$;
  \STATE For all $\lj\in\nd$: set $\ca{\lj}'\leftarrow\ca{\lj}$;
  \STATE Repeat the following until $Y_{\li}$ does not change for all $\li\in\sr$:
  \STATE~~~~~~~Run CSA with user set $\ur'$ and capacities $\{\ca{\lj}':\lj\in\nd\}$;
  \STATE~~~~~~~Let $X$ be the output service placement from CSA;
  \STATE~~~~~~~Set $\ur'\leftarrow\{\lk\in\ur':\se{\ui{\lk}}\cap\us{\lk}=\emptyset\}$;
  \STATE~~~~~~~For all $\lj\in\nd$: set $\ca{\lj}'\leftarrow\ca{\lj}'-\sum_{\li\in\sr:\lj\in\se{\li}}\sz{\li}$;
  \STATE~~~~~~~For all $\li\in\sr$: set $Y_{\li}\leftarrow Y_{\li}\cup X_{\li}$;
  \STATE Output service placement $Y$;
\end{algorithmic}
\end{algorithm}

In essence, RSA runs CSA and then, with the remaining users and remaining space on the nodes, runs CSA again to place new services in addition to those placed originally. It keeps repeating this with the remaining users and remaining space on the nodes until the combined service placement does not change. Because SA1 and SA2 respectively guarantee feasibility, CSA and RSA also guarantee feasibility. 

We can easily see that the approximation ratio of CSA and RSA is 
$\max\left\{ 1-e^{-\left(1-\sqrt{\fb}\right)^2} ; \left(1-e^{-1}\right) / 4 \right\}$,
because CSA adaptively chooses between SA1 and SA2 according to the one that gives the better approximation ratio, and the repeating step in RSA can only increase the reward which does not make the approximation ratio worse.

\subsection{Computational Complexity}
We first derive the time complexity of SA1/SA2/CSA.

The construction of the sets $\pa{\lj}{\ld}$ (for all $\ld\in\mathbb{N}\cup\{\bsl,\ssl\}$) takes a total time of $\mathcal{O}(|\sr|\cdot|\nd|)$ since, for every  pair $(\li,\lj)\in\sr\times\nd$, determining $\ld$ such that $\li\in\pa{\lj}{\ld}$ takes constant time.

Algorithm \ref{A3} takes a time of $\mathcal{O}(|\ls|\cdot|\ur|))$ which means Algorithm \ref{A2} takes a time of $\mathcal{O}(|\ls|\cdot|\sr|\cdot|(|\ls|\cdot|\ur|)$ which, since $|\ls|\in\mathcal{O}(|\sr|\cdot|\nd|)$, is equal to $\tilde{\mathcal{O}}(|\sr|^{3}\cdot|\nd|^{2}\cdot|\ur|)$.

Algorithm \ref{A5} takes a time of $\mathcal{O}(|\sr|\cdot|\nd|+|\ur|\cdot|\nd|)=\mathcal{O}(|\ur|\cdot|\nd|)$, which means Algorithm \ref{A4} takes a time of $\mathcal{O}(|\ur|\cdot|\nd|^2)$.

For the LP step, we have an input size of $\mathcal{O}(|\ur|\cdot|\nd|+|\sr|\cdot|\nd|)=\mathcal{O}(|\ur|\cdot|\nd|)$ and $\mathcal{O}(|\sr|\cdot|\nd|+|\ur|)$ variables. Karmarkar's algorithm~\cite{Karmarkar1984} takes a time of $\tilde{\mathcal{O}}\left(\left(|\sr|\cdot|\nd|+|\ur|\right)^{3.5}\left(|\ur|\cdot|\nd|\right)^2\right)$, which is equal to: 
\begin{equation}
\tilde{\mathcal{O}}\left(\left(|\sr|^{3.5}\cdot|\nd|^{5.5}\cdot|\ur|)+(|\nd|^2\cdot|\ur|^{4.5}\right)\right).
\label{eq:complexityCSA}
\end{equation}

Based on the above analysis, we can see that the bottleneck is the LP step. Hence the time complexity of SA1/SA2/CSA is given in (\ref{eq:complexityCSA}).
On each call to CSA in RSA, the number of served users increases by at least one. Hence, there are at most $|\ur|$ iterations. This gives the following complexity of solving SPSC with RSA:
\begin{equation}
\tilde{\mathcal{O}}\left(\left(|\sr|^{3.5}\cdot|\nd|^{5.5}\cdot|\ur|^2\right)+\left(|\nd|^2\cdot|\ur|^{5.5}\right)\right).
\end{equation}

The conversion from GSP to SPSC multiplies the number of users by $|\nd|$. Hence, the overall time complexity for converting GSP to SPSC and then solving the resulting SPSC (thus also solving the original GSP) using RSA has a complexity of:
\begin{align}
&  \tilde{\mathcal{O}}\left(\left(|\sr|^{3.5}\cdot|\nd|^{7.5}\cdot|\ur|^2)+(|\nd|^{7.5}\cdot|\ur|^{5.5}\right)\right).
\end{align}

\subsection{Bad Example of Greedy Algorithm}
\label{subsec:GreedyBadExample}

We consider a greedy algorithm that greedily places services starting with the highest reward. Such an algorithm has been shown to have a constant approximation ratio for the homogeneous setting with identical service sizes~\cite{TingICDCS2018}. In the heterogeneous setting we consider in this paper, the following example shows that it can have an arbitrarily bad approximation ratio for the GSP and SPSC problems.

Consider an example where we have $n+1$ services, $n+1$ users, and a single node with capacity $\ca{1} := 1$. Service $\li=1$ has size $\sz{1} := 1$, all the other services $\li > 1$ have size $\sz{\li}:= \frac{1}{n}$. Each user $\lk$ requires service $\ui{\lk} := \lk$. The reward of user $\lk=1$ is $\agf{1}(1) := 2$, and the reward of all the other users $\lk > 1$ is $\agf{\lk}(1) := 1$.
The greedy algorithm will simply place service $\li=1$ to the node, giving a reward of $2$. The optimal solution will place all the other $n$ services with $\li > 1$ on the node, giving a reward of $n$. This is true for all $n$. Hence, the approximation ratio of the greedy algorithm can be arbitrarily bad (close to zero) as $n \rightarrow \infty$ in this example.

\section{Simulation Results}
\label{sec:Simulation}

\begin{figure*}
    \centering
    \begin{subfigure}{0.23\textwidth}
        \centering
        \includegraphics[width=0.5\linewidth]{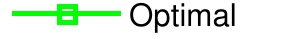}
    \end{subfigure}%
    ~
    \begin{subfigure}{0.23\textwidth}
        \centering
        \includegraphics[width=0.5\linewidth]{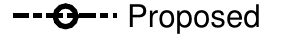}
    \end{subfigure}%
    ~
    \begin{subfigure}{0.23\textwidth}
        \centering
        \includegraphics[width=0.5\linewidth]{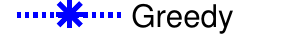}
    \end{subfigure}%
    ~
    \begin{subfigure}{0.23\textwidth}
        \centering
        \includegraphics[width=0.5\linewidth]{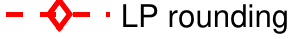}
    \end{subfigure}%
    
    \begin{subfigure}{0.19\textwidth}
        \centering
        \includegraphics[width=1\linewidth]{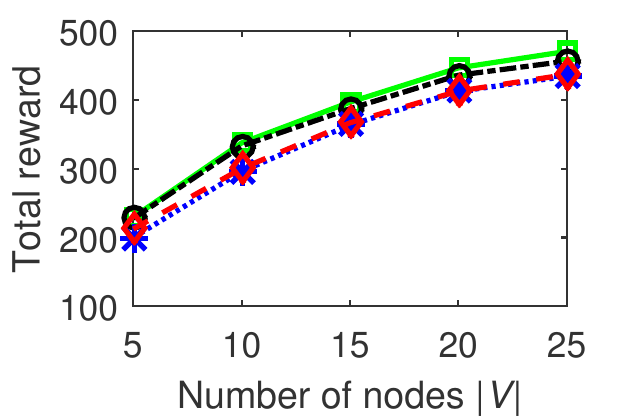}
    \end{subfigure}%
    ~
    \begin{subfigure}{0.19\textwidth}
        \centering
        \includegraphics[width=1\linewidth]{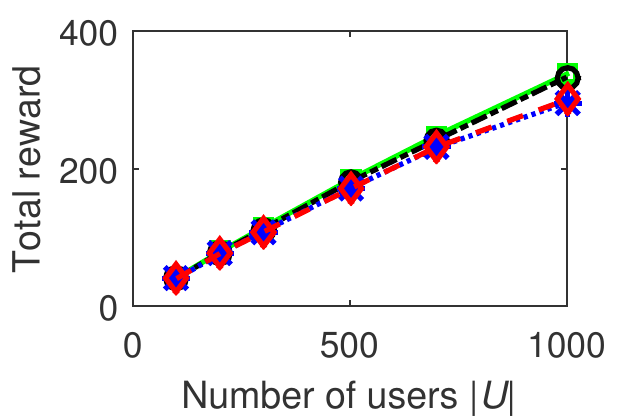}
    \end{subfigure}%
    ~
    \begin{subfigure}{0.19\textwidth}
        \centering
        \includegraphics[width=1\linewidth]{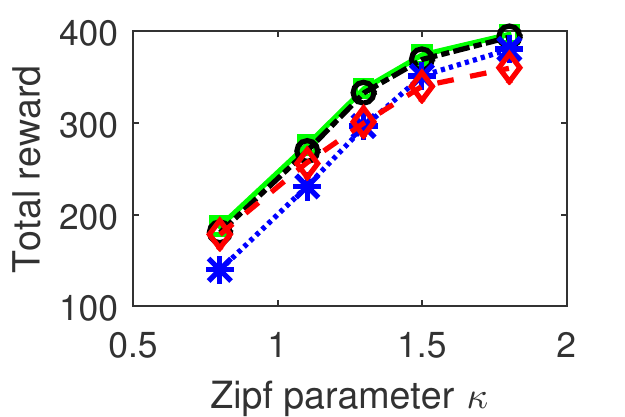}
    \end{subfigure}%
    ~
    \begin{subfigure}{0.19\textwidth}
        \centering
        \includegraphics[width=1\linewidth]{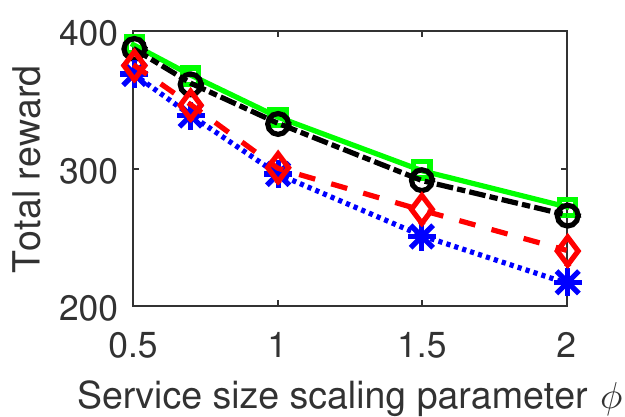}
    \end{subfigure}%
    ~
    \begin{subfigure}{0.19\textwidth}
        \centering
        \includegraphics[width=1\linewidth]{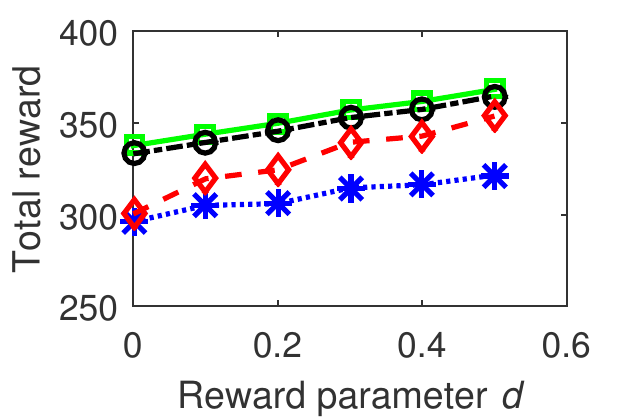}
    \end{subfigure}%
    \vspace{-0.2in}
    \begin{subfigure}{0.19\textwidth}
        \centering
        \includegraphics[width=1\linewidth]{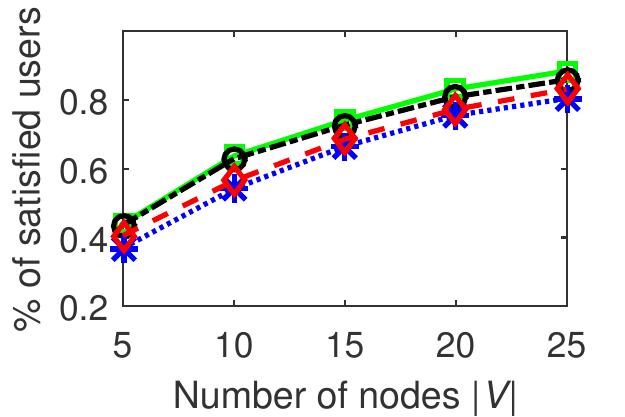}
    \end{subfigure}%
    ~
    \begin{subfigure}{0.19\textwidth}
        \centering
        \includegraphics[width=1\linewidth]{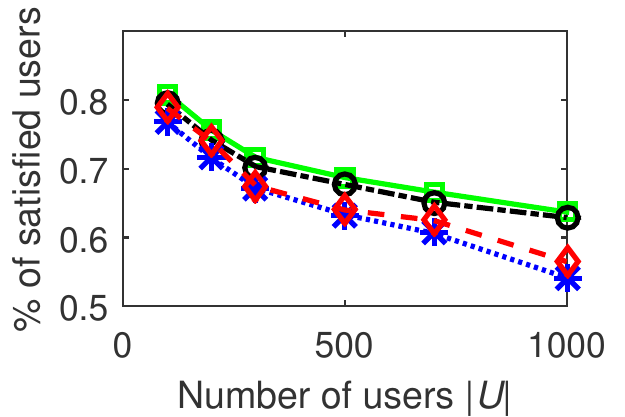}
    \end{subfigure}%
    ~
    \begin{subfigure}{0.19\textwidth}
        \centering
        \includegraphics[width=1\linewidth]{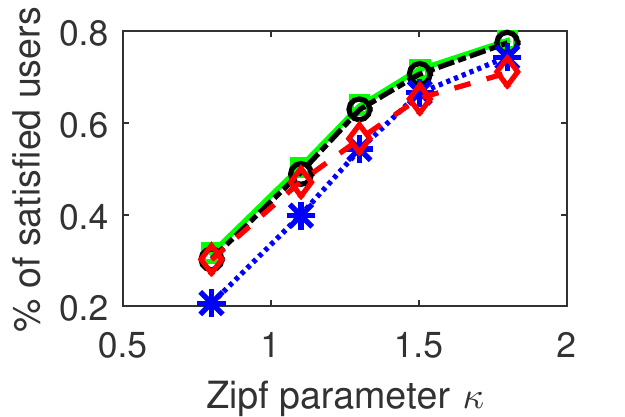}
    \end{subfigure}%
    ~
    \begin{subfigure}{0.19\textwidth}
        \centering
        \includegraphics[width=1\linewidth]{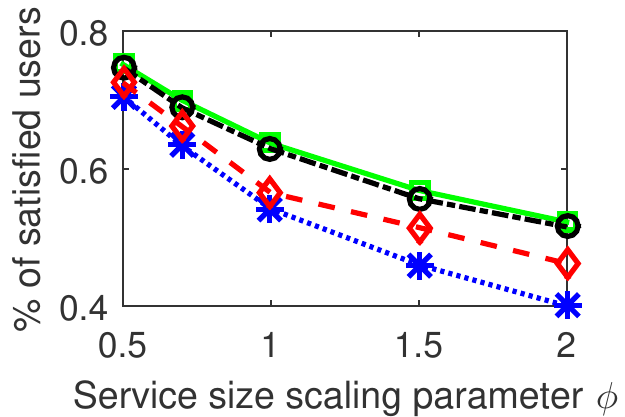}
    \end{subfigure}%
    ~
    \begin{subfigure}{0.19\textwidth}
        \centering
        \includegraphics[width=1\linewidth]{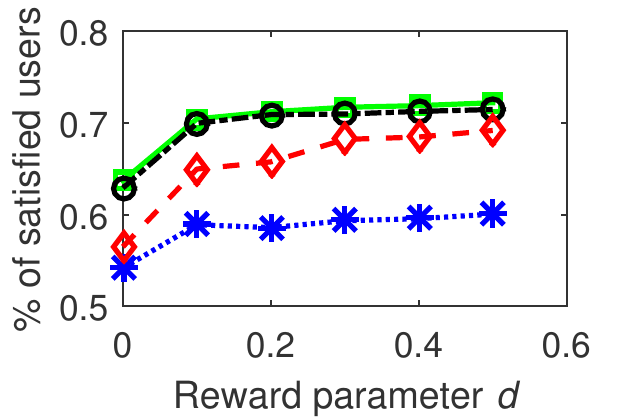}
    \end{subfigure}%
    
\caption{Simulation results with total reward and percentage of satisfied users (in SPSC) under different parameter settings.}
\label{fig:SimOthers}
\vspace{-0.2in}
\end{figure*}

We evaluate the performance of the proposed final algorithm, RSA (Algorithm~\ref{alg:RSA}), via simulations.
Inspired by practical measurements of mobile app popularity~\cite{Liu2015Mobidata} as well as related work~\cite{TingICDCS2018}, we assume that the service required by each user follows a Zipf distribution with parameter $\kappa$ and $|\sr|=1000$. The size of each service is $\phi\cdot(1 + Z_s / 14.13)$, where $Z_s$ is an exponentially distributed random variable with rate parameter $0.12$ and $\phi$ is a scaling parameter. These values are estimated from the results given in~\cite{Liu2015Mobidata} for the $33\%$ largest apps and normalized by the lower bound, because we consider that only large enough services need to be offloaded to other edge nodes, and the size of edge services can be on a different scale compared to mobile apps thus we do not directly use the mobile app sizes. The size of each node is randomly chosen from $\{4, 8, 16, 32\}$, to take into account that capacities of computational devices are usually integer powers of $2$. 

We randomly partition all nodes into two subsets, each service can either run on the first subset of nodes, the second subset of nodes, or both, with equal probability, to represent platform dependency of services. If user $\lk$ requires a service that cannot be run on node $\lj$, then we set the reward $\agf{\lk}(j)$ to zero.
Otherwise, we set $\agf{\lk}(j)$ as $Z_u + Z_n$, where $Z_u$ follows a uniform distribution within $[0.01, 1]$ to represent the importance of the service to the user, $Z_n$ follows a uniform distribution within $[-d, d]$ to represent the difference in rewards at different nodes. We also enforce that $Z_u + Z_n \in [0.01, 1]$ and re-sample the values of $Z_u$ and $Z_n$ until this is satisfied. 
We convert the GSP problem to its equivalent SPSC problem and present results for the SPSC problem.

We compare the proposed RSA algorithm (Algorithm~\ref{alg:RSA}) with the following baselines: 1) the optimal solution that has exponential time complexity; 2) the greedy algorithm proposed in~\cite{TingICDCS2018}; 3) an LP rounding mechanism that randomly rounds the solution $\{\lp{\li}{\lj}\}$ in (\ref{eq:LinearProgram}) to integers (with $\{\lp{\li}{\lj}\}$ as probabilities) until the node capacity has reached.

By default, we choose $|\ur| = 1000$, $|\nd| = 10$, $\phi = 1$, $\kappa = 1.3$, $d=0$. We vary one parameter while keeping the others fixed at the default value.
The average results of $10$ different simulation runs are shown in Fig.~\ref{fig:SimOthers}. 
In addition to the total reward, which is the objective we consider in this paper, we also plot the percentage of satisfied users (i.e., users that provide non-zero reward to the system) for comparison. 
We see that in all cases, both the total reward and the percentage of satisfied users of our proposed algorithm perform very close to the optimal, and better than the greedy and LP rounding baselines. This aligns with our theoretical result that shows our algorithm gives a good approximation ratio.

\section{Related Work}
\label{sec:RelatedWork}

In addition to the MEC-related work mentioned in Section~\ref{sec:introduction}, our work is also related to virtual network embedding (VNE) \cite{FischerVNESurvey}, which studies the scheduling of distributed tasks to multiple machines. However, service components that are shareable among multiple users is usually not considered in VNE, which is an important characteristic in MEC~\cite{TingICDCS2018}. 

Our problem bears some similarities with the data caching problem. However, a fundamental difference between data files and service programs is that data files can be partitioned in arbitrary ways without affecting the cache efficiency. Therefore, existing work on data caching usually considers identically sized files/contents~\cite{ShanmugamFemtoCaching,ShuklaJSAC2018}, which is inadequate for the placement of service programs.
Mathematically, the data caching problem has been extended to consider non-partitionable files of different sizes in~\cite{baev2008approximation}, which requires that the cost (opposite of the reward) is related to distances between nodes defined on a metric space. Such a distance metric is also a common assumption in facility location problems~\cite{vazirani2013approximation}. As discussed in~\cite{wang2007towards}, network delays often do not satisfy the triangle inequality, thus approaches based on metric assumptions~\cite{baev2008approximation,vazirani2013approximation} are not practical. The generalized assignment assignment problem~\cite{shmoys1993approximation} does not have metric assumptions, but it does not allow the replication of a service (item) into multiple copies that are placed in multiple edge nodes (bins).

Our problem is a special case of the $k$-column sparse packing problem studied in~\cite{bansal2012solving}, where a randomized approximation algorithm is proposed. Applying the approach in~\cite{bansal2012solving} to our case, one can get an algorithm with an approximation ratio of $(1-e^{-1})/7.25$
\if\citetechreport1
    (see \cite[Appendix~\ref{append:KSparse}]{techReportThisPaper} for details). 
\else
    (see Appendix~\ref{append:KSparse} for details). 
\fi
We improve on this result in two ways. First, we have a deterministic algorithm, which is more preferred due to its predictable behavior. Second, we have an approximation ratio that is at least $(1-e^{-1})/4$, which is better.

Our problem is also a special case of the separable assignment problem (SAP)
\cite{fleischer2011tight}, by considering users as items and nodes as bins in SAP. A subset of items (users) is feasible for a bin if the sum size of all services requested by these users is within the capacity of the node, where there is only one service instance for multiple users sharing the same service. Two approximation algorithms for SAP are proposed in~\cite{fleischer2011tight}. The first one has an approximation ratio of $1-e^{-1}$ and uses the ellipsoid method which has, when the number of users is much larger than the number of nodes, a higher complexity bound than our method (when solved with an efficient LP-solver)~\cite{CLP}. The ellipsoid method has also shown to be much slower than other LP-solvers in practice. Note that the approximation ratio of our approach also tends to $1-e^{-1}$ as $\beta$ tends to zero.
The second algorithm in~\cite{fleischer2011tight} gives an approximation ratio (upper bound) of $\frac{1}{2}$ with lower complexity than the ellipsoid method. Compared to this, our approach has a better approximation ratio when $\beta$ is small enough such that $1-e^{-\left(1-\sqrt{\fb}\right)^2} > \frac{1}{2}$, whereas we do not perform better when $\beta$ is large. Nevertheless, our work is based on a completely different algorithmic technique compared to~\cite{fleischer2011tight}.
Detailed investigation of the advantages and disadvantages of both approaches is left for future work.

\section{Conclusion}
\label{sec:Conclusion}

We have proposed an approximation algorithm for solving the service placement problem in MEC systems with heterogeneous service/node sizes and rewards. The algorithm has a constant approximation ratio, and has been shown to perform very close to optimum and better than baseline approaches in simulations. The algorithm includes a novel construction of slots on nodes. The design of the algorithm and the proof of the approximation guarantee is based on a highly non-trivial application of the method of conditional expectations.

\bibliographystyle{IEEEtran}
\bibliography{main}

\if\citetechreport1
    \clearpage
\fi

\appendices

\section{Proof of Theorem~\ref{theorem:NPHardnessGSP} (NP-Hardness)}
\label{append:NPHardProof}

Consider the decision version of the set cover problem (which is NP-complete~\cite{karp1972reducibility}):
We have a set $Z$, a set $\mathcal{A}$ of subsets of $Z$, and a number $c\in\mathbb{N}$.
The goal is to determine if a subset $\mathcal{B}$, of $\mathcal{A}$, exists that satisfies $|\mathcal{B}|\leq c$ and $\bigcup\mathcal{B}=Z$.

We define the following SPSC problem:
We have $|\mathcal{A}|-c+1$ services. One of these is special and denoted $\li'$. Let $\sr':=\sr\setminus\{\li'\}$.
We define $\nd:=\mathcal{A}$. So every node is a subset of $Z$.
For every $\li\in\sr'$ we have a user $\lk_{\li}$. For such a user we define $\ui{\lk_{\li}}:=\li$ and $\us{\lk_{\li}}=:\nd$.
For every $z\in Z$ we have a user $\lk'_{z}$. For such a user we define $\ui{\lk'_{z}}:=\li'$ and $\us{\lk'_{z}}:=\{Y\in\mathcal{A}:z\in Y\}$.
Every service $\li$ has size $\sz{\li}:=1$.
Every node $\lj$ has capacity $\ca{\lj}:=1$.
Every user $\lk$ has reward $\uw{\lk}:=1$.

If there exists a solution, $\mathcal{B}$, to the set cover problem,
we construct the following service placement $\rp$:
For each service $\li\in\sr'$ choose an node $\lj_{\li}$ in $\mathcal{A}\setminus\mathcal{B}$ in such a way that $\lj_{\li}\neq\lj_{\li^*}$ for $\li^*\in\sr'$ with $\li^*\neq\li$. This can indeed be done as $|\mathcal{A}\setminus\mathcal{B}|\geq|\mathcal{A}|-c=|\sr'|$.
For each service $\li\in\sr'$ define $\se{\li}:=\{\lj_{\li}\}$.
Define $\se{\li'}=\mathcal{B}$.
It is clear that this service placement is feasible as there is at most one service placed on each node. The fact that every user is satisfied can be seen as follows:
Given a service $\li\in\sr'$, the user $\lk_{\li}$ is satisfied as $\se{\ui{\lk_{\li}}}\cap\us{\lk_{\li}}=\se{\li}\cap\nd=\se{\li}\neq\emptyset$.
Given an element $z\in Z$, the user $\lk'_{z}$ is satisfied as there exists a set $Y'\in\mathcal{B}$ with $z\in Y$ and hence $\se{\ui{\lk'_z}}\cap\us{\lk'_z}=\se{\li'}\cap \{Y\in\mathcal{A}:z\in Y\}=\mathcal{B}\cap\{Y\in\mathcal{A}:z\in Y\}\supseteq\{Y'\}$.
This shows that there exists a solution to the above-defined SPSC in which every user is satisfied.

If there exists a solution to the above-defined SPSC in which every user is satisfied, 
we know that for every $\li\in\sr'$, the user $\lk_{\li}$ is satisfied, so $\ui{\lk_{\li}}=\li$ must be placed on some node. Let $\mathcal{C}$ be the set of all nodes $\lj$ on which a service in $\sr'$ is placed. Since only one service can be placed on each node, we have $|\mathcal{C}|\geq|\sr'|$, which is bounded below by $|\mathcal{A}|-c=|\nd|-c$.
Define $\mathcal{B}=\nd\setminus\mathcal{C}$ which, by above, has cardinality at most $c$.  For any $z\in Z$, we know that the user $\lk'_z$ is satisfied so there must be some node in $\us{\lk'_z}=\{Y\in \mathcal{A}:z\in Y\}$ on which service $\ui{\lk'_z}=\li'$ is placed. Let this node be denoted by $Y'$. Note that $Y'\notin\mathcal{C}$, because $\li'$ is placed there (thus, as only one service can fit on each node, no node in $\sr'$ can be placed there). This implies $Y'\in\mathcal{B}$ so $z\in\bigcup\mathcal{B}$. Since this holds for any $z\in Z$, we have that $\bigcup\mathcal{B}=Z$, showing that there is a solution, $\mathcal{B}$, to the set cover problem.

The above has shown that SPSC is NP-hard. Now, note that SPSC is an instance of GSP when, for every user $\lk\in\ur$, there exists some fixed $\uw{\lk}\in\mathbb{R}^+$ and $\us{\lk}\subseteq\nd$, such that $\agf{\lk}(\lj)=\uw{\lk}\cdot\id{\lj\in\us{\lk}}$. Hence, GSP is also NP-hard.

\section{Required lemmas}
\label{append:Lemmas}
Our proofs will require the following lemmas.

\begin{lemma}\label{p5}
Suppose we have $T$ independent draws in which the probability of an event $E$ happening on the $t$-th draw is $\rho_t$. Then the probability of event $E$ happening on either of the draws is equal to:
$$1-\prod_{t=1}^T(1-\rho_t)$$
\end{lemma}

\begin{prf}
This is a straightforward result in elementary probability theory.
\end{prf}

\begin{lemma}\label{p1}
Suppose we have $T$ independent draws in which the probability of an event $E$ happening on the $t$-th draw is $\rho_t$. Then the probability of event $E$ happening on either of the draws is at least: $$1-\exp\left(-\sum_{t=1}^T\rho_t\right)$$
\end{lemma}

\begin{prf}
Let $\Phi := \{ t\in\{1,2,...,T\}: \rho_t < 1\}$.
\if\citetechreport1
From elementary bounds of logarithmic function, we know that
\else
From \cite{topsok2006some}, we know that 
\fi
for any $t \in \Phi$, we have $\log(1 - \rho_t) \leq -\rho_t$.
Thus,
$$
\log\left(\prod_{t\in\Phi}(1 - \rho_t) \right)  = \sum_{t\in\Phi} \log(1 - \rho_t) \leq -\sum_{t\in\Phi} \rho_t.
$$
Taking exponentiation on both sides, we get
$$
\prod_{t\in\Phi} (1 - \rho_t) \leq \exp\left(-\sum_{t\in\Phi}\rho_t\right).
$$
For any $t\notin \Phi$, where $\rho_t = 1$, it is easy to see that $1-\rho_t = 0 \leq e^{-1} = \exp(-\rho_t)$.
Multiplying with the above and rearranging the inequality, we get
$$
1 - \prod_{t=1}^T (1 - \rho_t) \geq 1-\exp\left(-\sum_{t=1}^T\rho_t\right).
$$
The result then follows from Lemma \ref{p5}.

\end{prf}

\begin{lemma}\label{p2}
For any $a>0$ and $x\in[0,1]$ we have: $$1-e^{-ax}\geq x(1-e^{-a})$$
\end{lemma}

\begin{prf}
Let
$$g(a) := 1-e^{-ax} - x(1-e^{-a}) $$
Taking the first-order derivative of $g(a)$, we get
$$ g'(a)  = x (e^{-ax} - e^{-a})  \geq 0 $$
where the last inequality is because $ax \leq a$ due to $x\in [0,1]$ and $a>0$, thus $e^{-ax} \geq e^{-a}$. Therefore, $g(a)$ is non-decreasing with $a$ for any $a>0$. Combining this with the fact that $g(0) = 0$, we have $g(a) \geq 0$ and the result follows immediately.
\end{prf}

\begin{lemma}\label{p3}
Suppose we have $T$ independent draws in which the probability of an event $E$ happening on the $t$-th draw is $a\rho_t$ for some $a$. Then the probability of event $E$ happening on either of the draws is at least: $$(1-e^{-a})\min\left(1~,~ \sum_{t=1}^T\rho_t\right) $$
\end{lemma}

\begin{prf}
Let $x:=\sum_{t=1}^T\rho_t$. From Lemma \ref{p1} we have that the probability of event $E$ happening on either one of the draws is at least: $$1-\exp\left(-\sum_{t=1}^Ta\rho_t\right)=1-e^{-ax}.$$ If $x\leq 1$ then by Lemma \ref{p2} this is bounded below by $x(1-e^{-a})$ and if $x\geq 1$ this is bounded below by $(1-e^{-a})$. The result follows.
\end{prf}

\begin{lemma}\label{p4}
Suppose we have $T$ independent draws in which the probability of an event $E$ happening on the $t$-th draw is $a(1-\exp(-\rho_t))$ for some $a\leq 1$. Then the probability of event $E$ happening on either of the draws is at least: $$a\left(1-\exp\left(-\sum_{t=1}^T\rho_t\right)\right)$$
\end{lemma}

\begin{prf}
For $t'\leq T$ let $x_{t'}$ be the probability that event $E$ happens on either one of the first $t$ draws. Also, let $\rho'_{t'}:=\sum_{t=1}^{t'}\rho_{t}$. We maintain the inductive hypothesis (over $t$) that $x_t\geq a\exp(-\rho'_{t})$.

The inductive hypothesis clearly holds for $t=1$ as $x_1$ is the probability that $E$ happens in the first draw, by definition equal to $a\exp(-\rho_{1})=a\exp(-\rho'_{1})$.

Now suppose the inductive hypothesis holds for some $t$. Then we have:
\begin{align}
x_{t+1}&=x_t+(1-x_t)\left(a\left(1-e^{-\rho_{t+1}}\right)\right)\\
&=1-(1-x_t)\left(1-a\left(1-e^{-\rho_{t+1}}\right)\right)\\
&\geq1-\left(1-a\left(1-e^{-\rho'_t}\right)\right)\left(1-a\left(1-e^{-\rho_{t+1}}\right)\right)\\
&=a\left(1-e^{-\rho'_t}\right)+\left(1-a\left(1-e^{-\rho'_t}\right)\right)a\left(1-e^{-\rho_{t+1}}\right)\\
&\geq a\left(1-e^{-\rho'_t}\right)+\left(1-\left(1-e^{-\rho'_t}\right)\right)a\left(1-e^{-\rho_{t+1}}\right)\\
&= a\left(1-e^{-\rho'_t}\right)+e^{-\rho'_t}a\left(1-e^{-\rho_{t+1}}\right)\\
&=a\left(1-e^{-\rho'_t}e^{-\rho_{t+1}}\right)\\
&=a\left(1-\exp(-(\rho'_t+\rho_{t+1}))\right)\\
&=a(1-\exp(-\rho'_{t+1}))
\end{align}
which proves the inductive hypothesis. Hence, the probability of event $E$ happening on either of the trials is equal to $x_T$ which, by the inductive hypothesis, is bounded below by $a(1-\exp(-\rho'_{T}))$. This proves the result.
\end{prf}

\section{Proof of Theorem~\ref{theorem:FeasibilitySA2}}
\label{append:ProofFeasibleSA2}

As in the proof of Theorem \ref{FSA} we have, for all $\lj\in\nd$:
$$\sum_{\li\in\sr}\sz{\li}\cdot\id{\lj\in\set{\sla}{\li}}\leq\sum_{\su\in\ls:\sv{\su}=\lj}\sz{\sla(\su)}$$
 We then have three cases:
 \begin{itemize}
 \item If $\map(\lj)=\ha$ then we have a single slot $\su$, with $\sv{\su}=\lj$. We have, by definition of a slot allocation, that $\sla(\su)\in\pa{\lj}{\bsl}$ so $\sz{\sla(\su)}\leq\ca{\lj}$.
 \item If $\map(\lj)=\hb$ then we have a two slots $\su$ and $\su'$, with $\sv{\su}=\lj$ and $\sv{\su'}=\lj$. We have, by definition of a slot allocation, that $\sla(\su),\sla(\su')\in\pa{\lj}{\ssl}$ so $\sz{\sla(\su)},\sz{\sla(\su')}\leq\ca{\lj}/2$ and hence $\sz{\sla(\su)}+\sz{\sla(\su')}\leq\ca{\lj}$
 
 \item If $\map(\lj)=\hc$ then the result follows from Theorem \ref{FSA}, using $\{\li\in\sr:\sz{\li}\leq\ca{\lj}\fb\}$ instead of $\sr$ in the proof, and noting that all services on $\lj$ are in the set $\{\li\in\sr:\sz{\li}\leq\ca{\lj}\fb\}$
 \end{itemize}
 In all cases we have $\sum_{\su\in\ls:\sv{\su}=\lj}\sz{\sla(\su)}\leq\ca{\lj}$ which proves the feasibility of $\rpt{\sla}$

\section{Proof of Theorem~\ref{ER2}}
\label{append:ProofExpectedLowerBoundSA2}

To prove Theorem~\ref{ER2}, we first introduce the following additional theorems.

 \begin{theorem}\label{dp1}
 For any node $\lj\in\nd$ we have: $$\vo{\lj}\geq\frac{\fd}{\pr{\map(\lj)=\hc}}$$
 \end{theorem}
 
 \begin{prf}

Note first that $\fd\de{\lj}{\bsl}+\fd\pq{\lj}=\fd(\de{\lj}{\bsl}+\pq{\lj})\leq\fd(\de{\lj}{\bsl}+\de{\lj}{\ssl})$ which is equal to $\fd\sum_{\li\in\sr:\sz{\li}>\fb\ca{\lj}}\lp{\li}{\lj}$ and hence bounded above by $(\fd/\fb\ca{\lj})\sum_{\li\in\sr:\sz{\li}>\fb\ca{\lj}}\sz{\li}\lp{\li}{\lj}$ which, since $\fd=\fb=1/4$, is equal to $(1/\ca{\lj})\sum_{\li\in\sr:\sz{\li}>\fb\ca{\lj}}\sz{\li}\lp{\li}{\lj}$. We then have:
\begin{align}
\ca{\lj}&\geq\sum_{\li\in\sr}\sz{\li}\lp{\li}{\lj}\\
&=\sum_{\li\in\sr:\sz{\li}>\fb\ca{\lj}}\sz{\li}\lp{\li}{\lj}+\sum_{\li\in\sr:\sz{\li}\leq\fb\ca{\lj}}\sz{\li}\lp{\li}{\lj}\\
&\geq\ca{\lj}(\fd\de{\lj}{\bsl}+\fd\pq{\lj})+\sum_{\li\in\sr:\sz{\li}\leq\fb\ca{\lj}}\sz{\li}\lp{\li}{\lj}
\end{align}
which gives us
\begin{align}
\sum_{\li\in\sr:\sz{\li}\leq\fb\ca{\lj}}\sz{\li}\lp{\li}{\lj}&\leq\ca{\lj}(1-(\fd\de{\lj}{\bsl}+\fd\pq{\lj}))\\
\label{eq33}&=\ca{\lj}\pr{\map(\lj)=3}
\end{align}
Plugging this into the definition of $\vo{\lj}$ i.e.: $$\vo{\lj}:=\frac{\fd\ca{\lj}}{\sum_{\li\in\sr:\sz{\li}\leq\ca{\lj}\fb}\sz{\li}\lp{\li}{\lj}}$$ gives us the result
 \end{prf}
 
 \begin{theorem}\label{dp2}
For all nodes $\lj\in\nd$ we have $\fd/\pr{\map(\lj)=\hc}\leq 1$
 \end{theorem}
 
 \begin{prf}
 We now bound $\pr{\map(\lj)=\hc}$ below by $\fd$. We first note that $\pq{\lj}\leq 2$. this can be seen because if $\de{\lj}{\ssl}<2$ then $\pq{\lj}=\de{\lj}{\ssl}<2$ and if $\de{\lj}{\ssl}\geq 2$ then:
\begin{align}
\pq{\lj}&=\frac{1}{2}\de{\lj}{\ssl}\\
&=\frac{1}{2}\sum_{\li\in\pa{\lj}{\ssl}}\lp{\li}{\lj}\\
&\leq\frac{1}{2}\sum_{\li\in\sr:\sz{\li}\geq\fb\ca{\lj}}\lp{\li}{\lj}\\
&\leq\frac{1}{2\fb\ca{\lj}}\sum_{\li\in\sr:\sz{\li}\geq\fb\ca{\lj}}\sz{\li}\lp{\li}{\lj}\\
&\leq\frac{1}{2\fb\ca{\lj}}\sum_{\li\in\sr}\sz{\li}\lp{\li}{\lj}\\
&\leq\frac{1}{2\fb}\\
&=2
\end{align}
so in either case we have $\pq{\lj}\leq 2$. Note also that $\de{\lj}{\ssl}\geq\pq{\lj}$ so:
\begin{align}
\ca{\lj}&\geq\sum_{\li\in\sr}\sz{\li}\lp{\li}{\lj}\\
&\geq\sum_{\li\in\pa{\lj}{\bsl}}\sz{\li}\lp{\li}{\lj}+\sum_{\li\in\pa{\lj}{\ssl}}\sz{\li}\lp{\li}{\lj}\\
&\geq\frac{\ca{\lj}}{2}\sum_{\li\in\pa{\lj}{\bsl}}\lp{\li}{\lj}+\frac{\ca{\lj}}{4}\sum_{\li\in\pa{\lj}{\ssl}}\lp{\li}{\lj}\\
&=\frac{\ca{\lj}}{2}\de{\lj}{\bsl}+\frac{\ca{\lj}}{4}\de{\lj}{\ssl}\\
&\geq\frac{\ca{\lj}}{2}\de{\lj}{\bsl}+\frac{\ca{\lj}}{4}\pq{\lj}
\end{align}
and hence $\de{\lj}{\bsl}\leq 2-\pq{\lj}/2$. Putting together gives us:
\begin{align}
\pr{\map(\lj)=3}&=1-\fd(\de{\lj}{\bsl}+\pq{\lj})\\
&\geq1-\fd(2+\pq{\lj}/2)\\
&\geq 1-\fd(2+1)\\
&=1-3/4\\
\label{eq50}&=\fd
\end{align}
which implies the result.
 \end{prf}
 
 \begin{theorem}\label{PP}
For all services $\li\in\sr$ and all nodes $\lj\in\nd$, the probability that service $\li$ is placed on node $\lj$ (that is, $\pr{\lj\in\set{\sla}{\li}}$) is bounded below by $\fd(1-\exp(-\lp{\li}{\lj}))$.
\end{theorem}

\begin{prf}
First note that: $$\id{\lj\in\set{\sla}{\li}}=\id{\exists \su\in\ls:\sv{\su}=\lj~\operatorname{and}~\sla(\su)=\li}$$
We have five cases:
\begin{itemize}
\item $\sz{\li}>\ca{\lj}$. In this case we have $\lp{\li}{\lj}=0$ so the result holds trivially.
\item $\li\in\pa{\lj}{\bsl}$. In this case $\lj\in\set{\sla}{\li}$ only if $\map(\lj)=\ha$ which occurs with probability $\fd\de{\lj}{\bsl}$. Conditioned on the event that $\map(\lj)=\ha$ the probability that the single slot $\su$ with $\sv{\su}=\lj$ contains service $\li$ (i.e. $\sla(\su)=\li$) is equal to $\lp{\li}{\lj}/\de{\lj}{\bsl}$. Putting together we have $\pr{\lj\in\set{\sla}{\li}}=(\fd\de{\lj}{\bsl})(\lp{\li}{\lj}/\de{\lj}{\bsl})=\fd\lp{\li}{\lj}\geq\fd(1-e^{-\lp{\li}{\lj}})$ so the result holds.
\item $\li\in\pa{\lj}{\ssl}$ and $\de{\lj}{\ssl}<2$. In this case $\lj\in\set{\sla}{\li}$  only if $\map(\lj)=\hb$ which occurs with probability $\fd\pq{\lj}=\fd\de{\lj}{\ssl}$. Conditioned on the event that $\map(\lj)=\hb$ the probability that a given slot $\su$ with $\sv{\su}=\lj$ contains service $\li$ (i.e. $\sla(\su)=\li$) is equal to $\lp{\li}{\lj}/\de{\lj}{\ssl}$. Putting together we have $\pr{\lj\in\set{\sla}{\li}}\geq(\fd\de{\lj}{\ssl})(\lp{\li}{\lj}/\de{\lj}{\ssl})=\fd\lp{\li}{\lj}\geq\fd(1-e^{-\lp{\li}{\lj}})$ so the result holds.
\item $\li\in\pa{\lj}{\ssl}$ and $\de{\lj}{\ssl}\geq2$. In this case $\lj\in\set{\sla}{\li}$ only if $\map(\lj)=\hb$ which occurs with probability $\fd\pq{\lj}=\fd\de{\lj}{\ssl}/2$. Conditioned on the event that $\map(\lj)=\hb$ the probability that a given one of the two slots $\su$ with $\sv{\su}=\lj$ contains service $\li$ (i.e. $\sla(\su)=\li$) is equal to $\lp{\li}{\lj}/\de{\lj}{\ssl}$ so we have, by Lemma \ref{p3} with $a:=\lp{\li}{\lj}$, $T:=2$ and $\rho_t:=1/\de{\lj}{\ssl}$, that the probability that at least one of the two slots contains service $\li$ is bounded below by:
$$(1-e^{-\lp{\li}{\lj}})\min\left(1, 2/{\de{\lj}{\ssl}}\right)$$
which, since $\de{\lj}{\ssl}\geq 2$, is equal to 
$$(1-e^{-\lp{\li}{\lj}})2/{\de{\lj}{\ssl}}=(1-e^{-\lp{\li}{\lj}})/\pq{\lj}$$
 Putting together we have $\pr{\lj\in\set{\sla}{\li}}\geq(\fd\pq{\lj})(1-e^{-\lp{\li}{\lj}})/\pq{\lj}$ so the result holds.
 
\item $\li\in\pa{\lj}{\ld}$ for some $\ld\in\mathbb{N}$. In this case $\lj\in\set{\sla}{\li}$ only if $\map(\lj)=\hc$.  Conditioned on the event $\map(\lj)=\hc$ the probability that any given slot $\su$ with $\sv{\su}=\lj$ and $\sw{\su}=\ld$ contains service $\li$ (that is, $\pr{\lj\in\set{\sla}{\li}}$) is equal to $\lp{\li}{\lj}/\de{\lj}{\ld}$ so, as there are $\ns{\lj}{\ld}$ (which is at least $\vo{\lj}\de{\lj}{\ld}$) such slots we have, by Lemma \ref{p3} with $a:=\lp{\li}{\lj}$, $T:=\ns{\lj}{\ld}$ and $\rho_t:=1/\de{\lj}{\ld}$, that the probability that at least one such slot contains service $\li$ is bounded below by:
\begin{equation}\label{qe} (1-e^{-\lp{\li}{\lj}})\min\left(1, {\ns{\lj}{\ld}}/{\de{\lj}{\ld}}\right) \end{equation}
Noting that $\ns{\lj}{\ld}\geq \vo{\lj}\de{\lj}{\ld}$ and hence ${\ns{\lj}{\ld}}/{\de{\lj}{\ld}}\geq\vo{\lj}$ which, by Theorem \ref{dp1}, is bounded below by ${\fd}/{\pr{\map(\lj)=\hc}}$, Equation \ref{qe} is bounded below by:
\begin{equation} (1-e^{-\lp{\li}{\lj}})\min\left(1, {\fd}/{\pr{\map(\lj)=\hc}}\right) \end{equation} which, by Theorem \ref{dp2}, is equal to:
$$(1-e^{-\lp{\li}{\lj}}){\fd}/\pr{\map(\lj)=\hc}$$
 Putting together we have $\pr{\lj\in\set{\sla}{\li}}\geq\pr{\map(\lj)=\hc}(1-e^{-\lp{\li}{\lj}}){\fd}/\pr{\map(\lj)=\hc}$ so the result holds.
\end{itemize}
So in either case the probability that service $\li$ is placed on node $\lj$  is bounded below by $\fd(1-e^{-\lp{\li}{\lj}})$.
\end{prf}

\begin{prf}[Proof of Theorem~\ref{ER2}]
Too see this first note that:
$$\mathbb{E}\left(\sum_{\lk\in\ur}\sat{\sla}{\lk}\uw{\lk}\right)=\sum_{\lk\in\ur}\uw{\lk}\mathbb{E}(\sat{\sla}{\lk})$$ which, since $\sat{\sla}{\lk}$ is boolean, is equal to $$\sum_{\lk\in\ur}\uw{\lk}\pr{\sat{\sla}{\lk}=1}$$  
We shall now bound $\pr{\sat{\sla}{\lk}=1}$. Note first that $\sat{\sla}{\lk}=1$ if there exists a node $\lj\in\us{\lk}$ such that $\lj\in\set{\sla}{\ui{\lk}}$ so, since the existence of such a slot is independent for every node $\lj$ , we have, by Theorem \ref{PP} and Lemma \ref{p4} with $a:=\fd$ that:
$$\pr{\sat{\sla}{\lk}=1}\geq\fd\left(1-\exp\left(-\sum_{\lj\in\us{\lk}}\lp{{\ui{\lk}}}{\lj}\right)\right)$$
which is bounded below by $\fd(1-e^{-\lr{\lk}})$. Since $\lr{\lk}\leq 1$ this is, by Lemma \ref{p2}, bounded below by $\fd\lr{\lk}(1-e^{-1})$
Putting together gives us an expected total reward that is bounded below by:
$$\sum_{\lk\in\ur}\uw{\lk}\fd\lr{\lk}(1-e^{-1})=(1-e^{-1})\fd\arw$$
Theorem \ref{BOR} then gives us the result.
\end{prf}

\section{Proof of Theorem~\ref{ET}}
\label{append:ProofMethodOfCondExpectSA2}

We maintain the inductive hypothesis that $\exs{\pap}\geq\mathbb{E}\left(\sum_{\lk\in\ur}\sat{\sla}{\lk}\uw{\lk}\right)$ throughout the algorithm. 

Initially we have $\pap(\lj)=\emptyset$ for every $\lj\in\nd$ so $\mas{\psa}$ is the set of all possible construction maps and hence $\exs{\pap}=\mathbb{E}\left(\sum_{\lk\in\ur}\sat{\sla}{\lk}\uw{\lk}\right)$ so the inductive hypothesis holds.

Now suppose that the inductive hypothesis holds at some point during the algorithm. Note that $\{\mas{\ex{\map}{a}}:a\in\{\ha,\hb,\hc\}\}$ is a partition of $\mas{\pap}$ and hence:
\begin{align}
\exs{\ex{\map}{a'}}&=\max_{a\in\{\ha,\hb,\hc\}}\exs{\ex{\map}{a}}\\
&=\max_{a\in\ha,\hb,\hc}\mathbb{E}\left(\sum_{\lk\in\ur}\sat{\sla}{\lk}\uw{\lk}\Bigg|\map\in\mas{\ex{\map}{a}}\right)\\
&\geq~\mathbb{E}\left(\sum_{\lk\in\ur}\sat{\sla}{\lk}\uw{\lk}\Bigg|\map\in\mas{\ex{\map}{}}\right)\\
 &=\exs{\pap}
 \end{align}
which, by the inductive hypothesis, is bounded below by $\mathbb{E}\left(\sum_{\lk\in\ur}\sat{\sla}{\lk}\uw{\lk}\right)$. The fact that $\pap$ is then updated by $\ex{\map}{a'}$ then proves the inductive hypothesis.

The inductive hypothesis hence holds always which means, we have
$\mathbb{E}\left(\sum_{\lk\in\ur}\sat{\sla}{\lk}\uw{\lk}|\map=\fm\right)=\exs{\fm}
\geq\mathbb{E}\left(\sum_{\lk\in\ur}\sat{\sla}{\lk}\uw{\lk}\right)$ which proves the result.

\section{Proof of Theorem~\ref{theorem:CorrectnessOfExpectationSA2}}
\label{append:ProofCorrectnessOfExpectationSA2}

For all $\li\in\sr$ and $\lj\in\nd$ let $\eve{\li}{\lj}$ be the event that there exists some slot $\su$ with $\sla(\su)=\li$ and $\sv{\su}=\lj$.

Algorithm \ref{A5} first computes, for all $\li\in\sr$, $\lj\in\nd$ and $a\in\{\ha,\hb,\hc\}$:
$$\pbl{\li}{\lj}{a}:=\pr{\eve{\li}{\lj} | \map(\lj)=a}$$
via Lemma \ref{p5}. It then computes, by the law of total probability:
\begin{align}
\pbl{\li}{\lj}{\emptyset}&:=\pr{\eve{\li}{\lj}}\\
&=\pr{\exists{\su\in\ls}:\sla(\su)=\li, \sv{\su}=\lj}\\
&=\sum_{a\in\{\ha,\hb,\hc\}}\pr{\map(\lj)=a}\pbl{\li}{\lj}{a}
\end{align}

For all nodes $\lj\in\nd$, If $\ex{\map}{}(\lj)\neq\emptyset$ then $\pr{\eve{\li}{\lj} | \map\in\mas{\ex{\map}{}}}$ is equal to $\pr{\eve{\li}{\lj} | \map(\lj)=\ex{\map}{}(\lj)}$. On the other hand, if $\ex{\map}{}(\lj)=\emptyset$, then $\pr{\eve{\li}{\lj} | \map\in\mas{\ex{\map}{}}}$ is equal to $\pr{\eve{\li}{\lj}}$

So in either case we have that: $$\pr{\eve{\li}{\lj} | \map\in\mas{\ex{\map}{}}}=\pbl{\li}{\lj}{\ex{\map}{}(\lj)}$$

Algorithm \ref{A5} then computes, for all users $\lk\in\ur$, the quantity:
$$\pg{\lk}:=\pr{\sat{\sla}{\lk}=1|\map\in\mas{\ex{\map}{}}}$$
The fact that the algorithm computes the correct value of $\pg{\lk}$ can be seen as follows: $\sat{\sla}{\lk}=1$ if and only if there exists a $\lj\in\us{\lk}$ such that event $\eve{\ui{\lk}}{\lj}$ occurs. From Lemma \ref{p5} and the above we then have:
\begin{align}
\pg{\lk}&=1-\prod_{\lj\in\us{\lk}}(1-\pr{\eve{\ui{\lk}}{\lj}|\map\in\mas{\ex{\map}{}}})\\
&=1-\prod_{\lj\in\us{\lk}}(1-\pbl{\li}{\lj}{\ex{\map}{}(\lj)})
\end{align}
So we have shown that the algorithm correctly computes $\pg{\lk}$ for all users $\lk\in\ur$. The result then follows by noting:
\begin{align}
\exs{\ex{\map}{}}&=\mathbb{E}\left(\sum_{\lk\in\ur}\sat{\sla}{\lk}\uw{\lk}\Bigg|\map\in\mas{\ex{\map}{}}\right)\\
&=\sum_{\lk\in\ur}\uw{\lk}\mathbb{E}\left(\sat{\sla}{\lk}|\map\in\mas{\ex{\map}{}}\right)\\
&=\sum_{\lk\in\ur}\uw{\lk}\pr{\sat{\sla}{\lk}=1|\map\in\mas{\ex{\map}{}}}\\
&=\sum_{\lk\in\ur}\uw{\lk}\pg{\lk}
\end{align}

\section{Further Discussions on~\cite{bansal2012solving}}
\label{append:KSparse}
In~\cite{bansal2012solving} they give a randomized approximate algorithm for this problem when each element appears in at most $K$ of the knapsack constraints. Their algorithm has a parameter $\alpha$ and their approximation ratio (in expectation) is dependent on $\alpha$ and $K$. In the GSP case we have $K:=1$ and for $K=1$ the optimal tuning of $\alpha$ gives an approximation ratio of $(1-e^{-1})/7.25$. 
This is obtained by combining 
\cite[Lemma 3.9]{bansal2012solving} with \cite[Corollary 4.4]{bansal2012solving} and \cite[Lemma 4.2]{bansal2012solving},
and then maximizing over $\alpha$.
We improve on this result by having a deterministic algorithm with an approximation ratio of $(1-e^{-1})/4$.
Also in~\cite{bansal2012solving} they consider separately the case where the coefficients in the knapsack constraints are small compared to the minimum knapsack capacity. In the GSP case this means that the maximum service size is small compared to the minimum node capacity. However, for the case that $K=1$ the approximation ratio they prove is no better than the one for the general case whilst our approximation ratio is much better -- limiting to $(1-e^{-1})$ as the ratio of the services sizes to node capacities limits to zero.

\end{document}